\pdfoutput=1
\documentclass[journal,twocolumns]{IEEEtran}

\newcommand{\descr}[1]{\smallskip \noindent \textbf{#1}}

\usepackage{color}
\usepackage[hyphens]{url}
\usepackage{hyperref}
\usepackage[subtle]{savetrees}
\usepackage{fancyhdr}
\usepackage{bbm}
\usepackage{xspace}
\usepackage{enumerate}
\usepackage{lipsum}
\usepackage{multirow}
\usepackage[shortlabels]{enumitem}
\usepackage{amssymb}
\usepackage{makecell}
\usepackage{booktabs} 
\usepackage{amsmath}
\usepackage{mathbbol}
\usepackage{float}

\usepackage{microtype}
\usepackage[left=0.6in,right=0.6in,top=0.5in,bottom=0.5in,includehead,includefoot]{geometry}
\usepackage{graphicx}
\usepackage{subcaption}
\usepackage{mathtools}

\allowdisplaybreaks 
\usepackage{amsthm} 
\usepackage{graphicx}
\usepackage{comment}
\usepackage{xcolor}
\usepackage{alltt}
\usepackage{arydshln}
\usepackage{pifont}

\makeatletter

\usepackage{tikz}
\usetikzlibrary{decorations.pathreplacing,calc}

\usepackage{algorithm,algpseudocode}
\usepackage{setspace}

\newcommand{\sys}{\textsc{PrivTuna}\xspace}
\usepackage{tabularx}
\usepackage[T1]{fontenc}
\usepackage[utf8]{inputenc}
\usepackage{bm}
\definecolor{mygreen}{rgb}{0.0, 0.5, 0.0}
\DeclareCaptionType{copyrightbox}

\newif\ifcomment
\commenttrue 

\ifcomment
	\newcommand{\ap}[1]{\textbf{\em\color{blue}[AP: #1]}}

\else  
    \newcommand\ap[1]{}
\fi

\begin{document}

\title{How to Privately Tune Hyperparameters in Federated Learning? Insights from a Benchmark Study}

\author{
    \IEEEauthorblockN{Natalija Mitic\IEEEauthorrefmark{1}, Apostolos Pyrgelis\IEEEauthorrefmark{2}, Sinem Sav\IEEEauthorrefmark{3}\\ }
    \IEEEauthorblockA{\IEEEauthorrefmark{1}École Polytechnique Fédérale de Lausanne (EPFL)\\
    }
     \IEEEauthorblockA{\IEEEauthorrefmark{2}RISE Research Institutes of Sweden\\
    }
    \IEEEauthorblockA{\IEEEauthorrefmark{3}Bilkent University\\
    }
}

\maketitle
\thispagestyle{plain}
\pagestyle{plain}

\begin{abstract}
In this paper, we address the problem of privacy-preserving hyperparameter (HP) tuning for cross-silo federated learning (FL). We first perform a comprehensive measurement study that benchmarks various HP strategies suitable for FL. Our benchmarks show that the optimal parameters of the FL server, e.g., the learning rate, can be accurately and efficiently tuned based on the HPs found by each client on its local data. We demonstrate that HP averaging is suitable for iid settings, while density-based clustering can uncover the optimal set of parameters in non-iid ones. Then, to prevent information leakage from the exchange of the clients' local HPs, we design and implement \sys, a novel framework for privacy-preserving HP tuning using multiparty homomorphic encryption. We use \sys to implement privacy-preserving federated averaging and density-based clustering, and we experimentally evaluate its performance demonstrating its computation/communication efficiency and its precision in tuning hyperparameters.
\end{abstract}

\section{Introduction}\label{sec:intro}


Federated learning (FL) has become a promising approach for collaboratively training machine learning models on data originating from multiple clients (parties) without data outsourcing~\cite{Konency2016fed, federatedLearning1}. During the FL process, each client trains a local model on its sensitive data and communicates a model update with a central server that computes the global machine learning model by averaging the received updates. While recent research has exhaustively studied the convergence of the global model for various distributed learning algorithms~\cite{Konency2016fed,yu2019linear,Jiang2018,wang2019adaptive}, little effort has been devoted to tuning hyperparameters (HPs) which might affect the performance and accuracy of FL given its collaborative nature~\cite{kairouz2019,kuo2023noisy}. Traditional HP tuning techniques, such as grid/random search or Bayesian optimization, due to their iterative operation, introduce significant computation and communication overhead making them impractical for FL settings. Moreover, their tuning performance on heterogeneous, non-independent, and identically distributed (non-iid) data remains unclear.

Although the FL training process ensures that sensitive data does not leave the clients' local premises, recent research has shown that it does not provide strong levels of privacy as model updates leak information about the training data~\cite{Hitaj2017,Wang2019,Zhu2019,Melis2019,Nasr2019,Zhao2020,Jonas2020,Wainakh2022,Deng2021,Dimitrov2022}. To mitigate the risk of privacy leakage, researchers have proposed privacy-preserving federated learning (PPFL) solutions employing privacy-enhancing technologies (PETs) such as (i) differential privacy (DP)~\cite{shokri2015privacy,Nvidia_Fed,abadi2016deep,mcmahan2018LSTM,Wei2020,wu2019value}, (ii) secure multiparty computation (MPC)~\cite{zheng2019helen,Phong2017,Phong2018,Zhang20202,Bonawitz2016,Cock2021,wagh2019securenn,falcon,DTI,Chen,Zhu2020,jayaraman2018distributed,truex2019hybrid}, or (iii) homomorphic encryption (HE)~\cite{spindle,poseidon,rhode}. While all these solutions are crucial for preserving data privacy in FL, they assume that HPs, such as learning rate and momentum, are somehow predefined and configured. Performing HP tuning in the presence of PETs for FL becomes even more challenging given their modus operandi, e.g., DP solutions need to spend a significant portion of the privacy budget on the HP tuning itself, while MPC/HE approaches incur significant communication/computation overhead and thus cannot tolerate additional training iterations for HP tuning. As a result, there is a need for efficient HP tuning solutions for PPFL.


\begin{figure}[t]
\centering
    \centering
    \includegraphics[width=0.4\textwidth]{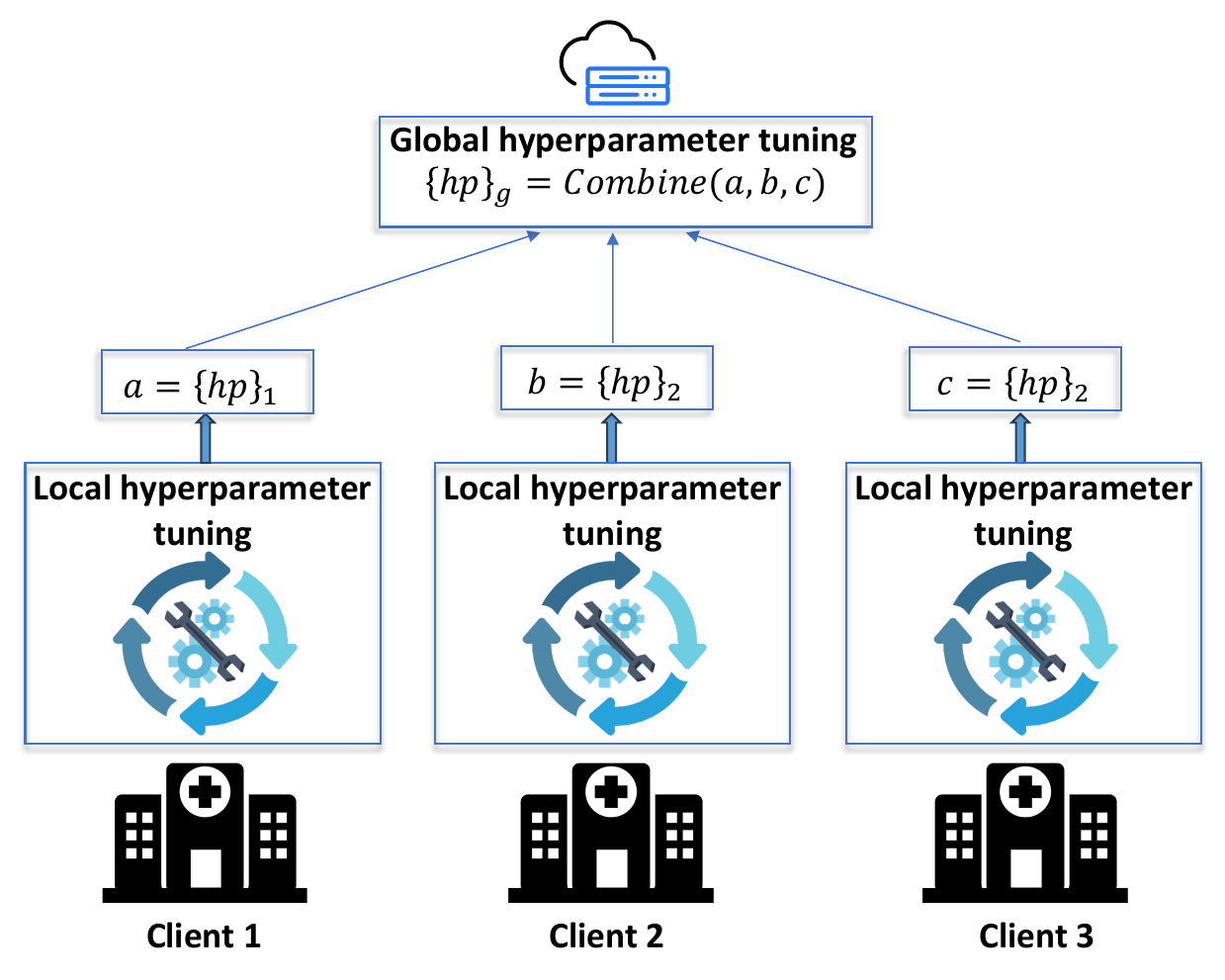} 
    \captionsetup{width=0.5\textwidth}
    \caption{Our HP tuning method for cross-silo federated learning. ${hp}_k$ and ${hp}_g$ denote the local best HP configuration at client $k$ and the global best configuration (denoted by $g$), respectively.}
    \label{fig:method}
\end{figure}

Existing \emph{efficient} HP tuning methods for FL, such as FLoRa~\cite{flora}, rely on the \textit{single-shot} HP tuning paradigm: Each client locally discovers its optimal set of parameters and communicates it to the central server that decides the most suitable HPs for the federated learning before the training process starts. However, sharing local hyperparameters with the FL server might result in new forms of privacy leakage as prior work has shown that data samples with special attributes, e.g., outliers, can noticeably skew the optimal HP configuration and remain vulnerable to privacy attacks such as membership inference~\cite{papernot2022hyperparameter}. While employing noise-based techniques during HP tuning is a potential solution~\cite{papernot2022hyperparameter,wang2023dphypo}, these might severely degrade the utility and the performance of the trained model, raising the urgency for efficient approaches that are both private and retain the accuracy of the federated hyperparameter tuning procedure.


In this work, we aim to address this gap by proposing an efficient and accurate method for tuning HPs in cross-silo FL by letting each client discover their optimal hyperparameters and by combining them in a privacy-preserving (PP) manner at the central server (Figure~\ref{fig:method}). To better understand how to combine the local hyperparameters found by each client at the server side, we first perform a comprehensive measurement study on various datasets and model architectures simulating both iid and non-iid FL settings, and we benchmark a wide range of tuning strategies. Inspired by aggregation methods commonly used in FL, we benchmark strategies such as computing the mean~\cite{federatedLearning1}, the median~\cite{yin2021byzantinerobust,chen2019distributed}, the trimmed mean~\cite{yin2021byzantinerobust}, trimmed mean/median with the highest validation accuracy, and density-based clustering~\cite{Agrawal2021,briggs2020federated}, and we compare their performance. Interestingly, our study shows that HP averaging is a simple and effective technique for iid settings, while density-based clustering can uncover the optimal hyperparameters in non-iid ones. Then, to protect the privacy of the clients' local HPs (hence, their sensitive data) we design a novel framework, \sys, that facilitates their privacy-preserving combination at the server using multi-party homomorphic encryption techniques. We use \sys to implement two privacy-preserving combination strategies for HP tuning, i.e., federated mean and density-based clustering, and we evaluate its performance in terms of computation/communication overhead and hyperparameter tuning precision.

The primary contributions of this paper are the following:

\begin{itemize}[leftmargin=*]

\item A comprehensive measurement study on various datasets, model architectures, and settings, that benchmarks various HP combination strategies on these settings and investigates the link between the hyperparameters of the clients and the FL server.

\item The identification of two efficient strategies for tuning the server hyperparameters based on the optimal HPs discovered by each client: (i) collective averaging for iid settings, and (ii) density-based clustering for non-iid, which outperform prior work following a similar paradigm~\cite{flora}.

\item The design of \sys, the first framework for privacy-preserving hyperparameter tuning for FL based on multi-party homomorphic encryption (MHE) techniques.

\item The evaluation of \sys on two privacy-preserving strategies, i.e., federated mean and density-based clustering, in terms of computation/communication overhead and hyperparameter tuning precision.




\end{itemize}
Our implementation and supplementary results can be found at \url{https://github.com/sinemsav/hyperparams/}.\\
\descr{Paper Organization.} The rest of the paper is organized as follows: In Section~\ref{sec:background}, we introduce the background on federated learning and hyperparameter tuning. We describe the problem under investigation and the roadmap of our solution in Section~\ref{sec:problem}. We detail the methodology, experimental setup, and results of our benchmarks in Section~\ref{sec:design}. We introduce \sys, a novel framework for privacy-preserving HP tuning in FL in Section~\ref{sec:privFL}. We review the related work in Section~\ref{sec:related} and conclude in Section~\ref{sec:conclusion}.


\section{Background}\label{sec:background}

We here provide the necessary background for our work. We first describe federated learning (FL) and its privacy-enhanced variants (Section~\ref{subsec:fl}) and then we discuss hyperparameter (HP) tuning for FL and the related challenges (Section~\ref{subsec:hyperparamtuning}).

\subsection{Federated Learning}\label{subsec:fl}

\begin{algorithm}[t]
\small
\caption{Federated Averaging Algorithm ($\textsf{FedAvg}$)}
\label{algo:fl}
\begin{algorithmic}[1]
\State Initialize the global model ${W}^0$ \Comment{Server executes}
\For{$k=0 \rightarrow e-1$} \Comment{Global iterations}
\State Choose $S$ random subset of $N$ clients
\State  Send ${W}^0$ to each data holder in $S$
\State \textbf{Each $S_i \in S$}
 \For{$\ell=0 \rightarrow l-1$} \Comment{Local iterations}
 \State  ${W}_{i}^{k+1} \leftarrow \textsf{LocalIteration}(\cdot)$ \Comment{at each $S_i \in S$}
 \EndFor
\State {\color{blue}${W}^{k+1} \leftarrow \sum_{i=1}^{N} \frac{n_i}{n} { W}_{i}^{k+1}$ \Comment{Server executes with its HPs}}
\EndFor
\end{algorithmic}
\end{algorithm}

Federated Learning (FL) is a collaborative approach to machine learning that allows multiple entities to collectively train a model without sharing their respective datasets~\cite{federatedLearning1,Konecny2016,Konency2016fed}. In FL, each data holder independently conducts several training iterations using their local data, and the resulting local models are combined into a global model through an aggregation server. While there exist various FL (aggregation) algorithms, in this work we focus on the most widely adopted one called $\textsf{FedAvg}$~\cite{federatedLearning1}; it is summarized in Algorithm~\ref{algo:fl}, with ${W}_{i}^{k}$ denoting the machine learning model weights of the $i^{th}$ client during the $k^{th}$ iteration ($i$ is omitted for the global model weights at the server). The model updates from each client $S_i$ are aggregated to the global model via weighted averaging (Line 9, Algorithm~\ref{algo:fl}, with $n_i$ the number of local samples at the $i^{th}$ client and $n$ the total number of samples). Note that the server utilizes its HPs, e.g., learning rate or momentum, to update the global model (Line 9, Algorithm~\ref{algo:fl}).

\descr{Privacy-Preserving Federated Learning.} Recent literature has proposed several gradient-targeting inference attacks~\cite{Hitaj2017,Wang2019,Zhu2019,Melis2019,Nasr2019,Zhao2020,Jonas2020,Jin2021,Wainakh2022,Deng2021,Dimitrov2022} which demonstrate that FL in its vanilla form does not provide strong privacy protection for the individual samples of the clients' datasets. Consequently, researchers propose to strengthen FL by incorporating privacy-enhancing technologies (PETs) in its training process, e.g., differential privacy, secure multiparty computation, and homomorphic encryption, to protect the gradients shared by the clients with the central server~\cite{zheng2019helen,spindle,poseidon,rhode,corrigan2017prio,Froelicher2021,Sav2022,flash,trident,shokri2015privacy,Nvidia_Fed,abadi2016deep,McMahan2018,Wei2020,wu2019value}. However, all these works assume that the FL hyperparameters are somehow predefined, which is an unrealistic assumption for practical applications. Moreover, using PETs to strengthen the privacy of FL, introduces additional costs, e.g., utility loss, computation and communication overhead, which prohibits adaptive solutions that perform HP tuning while training the global model (see Section~\ref{sec:related}). To this end, the main focus of our work is to propose efficient techniques for HP tuning in FL which take place before the federated training starts and which are suitable for its PP variants.

\subsection{HP Tuning for Federated Learning}\label{subsec:hyperparamtuning}

Machine learning algorithms operate using a set of configurable parameters, i.e., hyperparameters (HPs), which control various aspects of the learning process, such as the convergence criteria or the regularization strength, and overall govern the performance of the model. Examples of HPs are the learning rate, batch size, the regularization coefficient, and momentum. Inappropriate HP configurations can lead to poor model performance, such as underfitting, overfitting, and limited generalization capabilities. Determining the optimal combination of HPs, however, is a non-trivial task due to computational complexity. To this end, HP tuning refers to the process of searching for the best HP configuration that maximizes the model's performance on a given task. There are several well-known approaches for centralized HP tuning, such as manual search, grid search~\cite{LeCun2012,lecun2000,Bellman1961}, random search~\cite{bergstra12a}, Bayesian optimization~\cite{Mockus}, and genetic algorithms~\cite{Bardanet}. However, these approaches cannot be easily integrated into FL due to the high number of training iterations that they require. Thus, decentralized tuning for FL is an active research field with various proposals based on weight sharing~\cite{khodak2019weight,khodak2021federated}, neural architecture search~\cite{zhu2019multiobjective,Zhu2021Nas,he2021noniid} or Thompson sampling~\cite{Dai2020}. Nonetheless, key challenges such as computation and communication efficiency, as well as privacy, remain open for HP tuning in FL. We refer the reader to Section~\ref{sec:related} for further details.

In the measurement and benchmarking part of this work (Section~\ref{sec:design}), we rely on grid search for tuning the local HPs of each FL client and on its federated version for discovering the optimal server parameters. Grid search involves specifying a predefined set of values for each HP under consideration and exhaustively evaluating all possible combinations. The combination that achieves the best evaluation metric, e.g., validation accuracy, is then chosen for the task. We rely on grid search because (i) it guarantees coverage of the entire intended search space, and (ii) it is straightforward to implement. We here note that our main focus is not to choose the best HP technique; our method can be extended to other HP tuning methods. Overall, our aim is to discover efficient and accurate methods for combining the client local HPs in order to derive the optimal FL parameters at the server side.



\section{Problem Statement and Roadmap}\label{sec:problem}

We first discuss the envisioned setting and the problem statement (Section~\ref{subsec:problem}). Then, we provide a roadmap of our approach to tackle the problem under investigation (Section~\ref{subsec:roadmap}).

\subsection{Problem Statement}\label{subsec:problem}

We consider a cross-silo federated learning setting, where several (tens or hundreds of) entities (e.g., hospitals, companies, institutions) aim at collaboratively training a machine learning model on their sensitive data. To this end, they need to employ a privacy-preserving federated learning (PPFL) solution where the aggregator receives model updates enhanced with a privacy-enhancing technology, e.g., differential privacy, secure multi-party computation or homomorphic encryption (see Section~\ref{subsec:fl}). To ensure the convergence and performance of the trained model, the entities desire to tune any hyperparameters (HPs) related to the learning process \textbf{before} the federated training starts. The HP tuning process should be efficient, accurate, and should respect the privacy of each client's sensitive data.

We here note that federated learning requires the tuning of both server- and client-side hyperparameters. Server-side HPs are used by the server to update the global model (Line 9, Algorithm~\ref{algo:fl}), while client HPs are employed to perform local training iterations at the client side. In this work, we focus on the problem of configuring \textbf{server-side parameters}, e.g., server learning rate or momentum, as these parameters play a significant role in the accuracy of the global model, especially when PPFL solutions are employed. For instance, differentially private solutions introduce noise in the learning process, thus these parameters need to be adapted accordingly~\cite{papernot2022hyperparameter}, while secure multi-party computation and homomorphic encryption approaches require the approximation of non-linear functions (e.g., activations) whose performance is affected by these hyperparameters~\cite{spindle, poseidon, rhode, knott2021crypten, wagh2019securenn, keller2022secure, cryptoDL}.

\begin{algorithm}[h]
\small
\caption{Single-shot HP tuning for Cross-silo FL.}
\label{algo:method}
\begin{algorithmic}[1]
\For{Each client $S_i \in S$}
\State Perform local HP optimization (LHO) for all HPs in the set $\{hp\}$              \Comment{Client executes} 
\State Send the best or all ($\{hp\}_i$, $\{accuracy\}_i$) pairs found by LHO to the server
\EndFor
\State $\{hp\}_g = \textsf{Combine}(\{hp\}_i, \{accuracy\}_i)$  \Comment{Server executes} 
\end{algorithmic}
\end{algorithm}

\subsection{Roadmap}\label{subsec:roadmap}

To address the efficient, accurate, and privacy-preserving hyperparameter tuning problem for cross-silo federated learning, we make the following observation: As federated learning operates by aggregating models learned on each client's local dataset, the effectiveness of the global model should depend on the HPs that can be locally configured by each client. Inspired by this, we hypothesize that an HP configuration achieving good performance on the global federated model can be derived from the client local HPs. This, in turn, implies that HPs proven effective on individual client data are expected to be effective when combined together. As such, following the footprint of efficient single-shot solutions for hyperparameter tuning in federated learning~\cite{flora}, our objective is to infer the optimal global HPs solely based on the local HPs discovered by the clients. Algorithm~\ref{algo:method} presents the high-level overview of our approach for efficient HP tuning in cross-silo federated learning settings: (i) Each client performs local optimization (LHO) to tune the HPs in the set $\{hp\}$ on its dataset. Then, (ii) each client communicates the optimal sets of parameters along with an indicator of their performance (e.g., the accuracy achieved on a local validation set) to the server. Finally, (iii) the server combines ($\textsf{Combine}(\cdot)$, Line 5) the parameter sets received from all the clients to derive the optimal global parameters $\{hp\}_g$.


Aiming to find effective strategies to combine (i.e., $\textsf{Combine}(\cdot)$, Line 5, Algorithm~\ref{algo:method}) the client local hyperparameters and derive the optimal global hyperparameters at the server side, we perform a comprehensive measurement study to better understand the connection between local and global HPs in FL based on various factors such as the dataset, the model architecture, the client configuration, and the data distribution (Section~\ref{sec:design}). Through our measurement study, we benchmark various tuning strategies that infer server HPs from the client HPs. Inspired by aggregation methods commonly used in FL, we experiment with combination strategies such as computing the mean~\cite{federatedLearning1}, the median~\cite{yin2021byzantinerobust,chen2019distributed}, the trimmed mean~\cite{yin2021byzantinerobust}, trimmed mean/median with the highest validation accuracy, and density-based clustering~\cite{Agrawal2021,briggs2020federated}, and we compare their performance on various configurations. Then, to address the privacy concerns raised by sharing local HPs with the FL server, we introduce a novel framework, named \sys, that enables the HP combination at the server side in a privacy-preserving manner (Section~\ref{sec:privFL}). \sys is based on multi-party homomorphic encryption and can be used to develop various HP tuning techniques for FL with strong privacy guarantees. We use \sys to develop two HP combination strategies that our benchmarking indicated as promising approaches for cross-silo federated learning settings, and we evaluate their performance in terms of computation/communication overhead, and HP tuning precision.



\section{Benchmarking Hyperparameter Tuning Strategies for Federated Learning}\label{sec:design}

In this section, we perform a measurement study to better understand the relation between the client and server hyperparameters in various federated learning settings through benchmarking various HP tuning strategies. Following our solution roadmap (Section~\ref{sec:problem}), the aim of the measurement study is to discover suitable strategies for combining (i.e., $\textsf{Combine}(\{hp\}_i, \{accuracy\}_i)$, Line 5, Algorithm~\ref{algo:method}) the clients optimal HPs and to derive the appropriate server HPs. We first describe our methodology (Section~\ref{sec:methodology}), and then we discuss our experimental setup (Section~\ref{sec:exp-setup}) and the corresponding results (Section~\ref{sec:results}).

\subsection{Methodology}\label{sec:methodology}

To benchmark combination methods and find a suitable one for the global HP configuration based on the local HPs discovered by the clients, we conduct a measurement study on various datasets with different neural network architectures simulating both iid and non-iid federated learning settings. Our measurement study follows the subsequent footprint: (i) We first perform local HP optimization (\textbf{LHO}) per client to derive the optimal HPs on each client dataset. We also perform global HP optimization (\textbf{GHO}) using federated grid search to construct a \textit{ground truth} for the optimal server parameters. 
Then, (ii) following Algorithm~\ref{algo:method}, we employ various combination strategies to derive the server HPs based on the local HP optimization results of each client (i.e., their $\{hp\}_i$, $\{accuracy\}_i$ pairs). The tuning strategies that we benchmark for the $\textsf{Combine}(\cdot)$ method of Algorithm~\ref{algo:method} are the following:

\begin{itemize}[leftmargin=*]
    \item \textbf{Mean:} Inspired from vanilla $\textsf{FedAvg}$~\cite{federatedLearning1}, we compute the mean of all the HPs obtained by LHO execution at each client.

    \item \textbf{Median:} Based on the (coordinate-wise) median aggregation rule for robust federated learning~\cite{yin2021byzantinerobust,chen2019distributed}, we adopt a method that estimates the median of all the HPs obtained by LHO execution at each client.
    
    \item \textbf{Trimmed-mean:} We apply a similar strategy to the trimmed-mean aggregation rule for federated learning~\cite{yin2021byzantinerobust}. Specifically, we trim the largest and smallest $10\%$ of the sorted (by their values) HPs obtained through LHO at each client, and we compute the mean of the remaining HPs.

    \item \textbf{Mean/Median of Best HPs:} We sort the HPs of each client with respect to the validation accuracy they achieve. Then, for each client, we choose the top $5\%$ of HPs, and we compute their mean/median at the server side.
    
    \item \textbf{DBSCAN:} Following recent literature on clustered federated learning~\cite{briggs2020federated,Agrawal2021}, we use density-based clustering -- a method well-suited to low-dimensional data and robust to outliers (see Appendix~\ref{subsec:dbscan})-- to group the best HPs discovered by LHO at each client based on their validation accuracy. We compute the cluster centroids as the final HP set.

    
\end{itemize}

\noindent Finally, (iii) we compare the outcomes of each HP combination strategy with the ground truth baseline (i.e., the \textbf{GHO}), aiming to identify the most suitable methods for various settings.


\subsection{Experimental Setup}\label{sec:exp-setup}

We now describe the experimental setup of our measurement study, namely, the datasets that we employ (Section~\ref{sec:datasets}), the model architectures (Section~\ref{sec:models}), the federated learning settings that we simulate (Section~\ref{sec:flsettings}), and the HPs and their configurations (Section~\ref{sec:hyperparams}). All our experiments are implemented in Python, leveraging on the FedJax library~\cite{fedjax2021} for the FL operations. 

\subsubsection{Datasets}\label{sec:datasets}

We employ four datasets with varying input and output sizes, and different tasks, and we perform HP tuning based on validation accuracy which measures the model's ability to predict labels for a validation set that constitutes a subset of the original training data. We list the datasets below:

\descr{MNIST.} MNIST~\cite{MNIST} is a widely used database comprising $28 \times 28$ pixel grayscale images of handwritten digits, with 10 classes (from 0 to 9). The dataset contains $60$K training and $10$K testing samples.

\descr{Extended MNIST (EMNIST).} EMNIST~\cite{cohen2017emnist} is an extension of the MNIST dataset that contains a total of $\sim$814K $28 \times 28$ pixel grayscale images of handwritten characters including upper- and lower-case letters and digits. The number of classes is $62$.

\descr{Street View House Numbers (SVHN).} SVHN~\cite{svhn} is digit recognition dataset, obtained from real-world images of house numbers which were collected by Google Street View. The training and test datasets comprise approximately $100$K digits, with $32 \times 32$ pixel RGB images.

\descr{CIFAR-10.} CIFAR-10~\cite{cifarPaper} is a dataset of $60$K images labeled across 10 categories (e.g., cat, ship, dog, etc.). Each image in the dataset is a $32\times32$ pixel RGB image.

\subsubsection{Model Architectures}\label{sec:models}

In our study, we rely on 3 different neural network architectures:

\descr{Network A.} This model consists of two sequential convolutional layers with a kernel size $5 \times 5$, and 32 output channels followed by an average pooling (with window shape 2 and stride of 1), and a dense layer with a hidden layer of 64 neurons. The model has a total of $54,274$ parameters. We use Network A in all our experiments with the MNIST dataset.

\descr{Network B~\cite{li2021federated}.} This model consists of two sequential convolutional layers with kernel size $5 \times 5$, and 6 and 16 output channels, respectively. We use average pooling (with window shape 2 and stride of 1) and two dense layers of 120 and 84 neurons, respectively. We use Network B in all our experiments with EMNIST and SVHN dataset. The EMNIST and SVHN models have $29,896$ and $32,756$ parameters, respectively.

\descr{Network C.} The model consists of six sequential convolutional layers with kernel size $5 \times 5$ using average pooling (with window shape 2 and stride 1), and a dense layer with $128$ neurons. The model has a total of $699,018$ parameters. To prevent overfitting we use dropout, a regularization method that randomly omits units during the training process of a neural network. We use Network C in our experiments with the CIFAR-10 dataset.

\subsubsection{Federated Learning Settings}\label{sec:flsettings}

Our measurement study accounts for both iid and non-iid federated learning settings. To simulate the former we equally distribute the original dataset to $N$ clients with $N=[10, 20, 50]$. To simulate a more realistic FL deployment, we consider a non-iid environment with $N=[10, 20]$ clients. Non-iid refers to data that is not independently and identically distributed across samples. In other words, the data is not drawn from the same distribution and the samples are not independent of each other. We generate various non-iid distributions following the methods described in~\cite{li2021federated} to account for label, feature, and quantity skews:

\descr{Label Skew.} In label distribution skew, the label distributions vary across clients. Each client \(i\) is allocated a proportion \(p_{i}\) of the samples of each label according to a Dirichlet distribution ($\text{Dir}_N$), where \(p_{i} \sim \text{Dir}_N(\beta_{\ell})\). In our experiments, we use \(\beta_{\ell} \in \{0.1, 1.0, 5.0\}\).

\descr{Feature Skew.} In feature distribution skew, the feature distributions vary across clients. First, we partition randomly and equally the entire dataset among each client \(i\). Then, each client's local dataset is subjected to varying degrees of Gaussian noise \(n_i\) in order to generate distinct feature distributions, where \(n_i \sim \mathcal{N}(0,\,\frac{\beta_{f} * i}{N})\). In our evaluations, we use \(\beta_{f} \in \{0.02, 0.1\}\).

\descr{Quantity Skew.} In quantity skew, the size of the local dataset varies across clients. Each client \(i\) is allocated a proportion \(q_i\) of the total data samples according to a Dirichlet distribution, where \(q_i \sim \text{Dir}_N(\beta_{q})\). We use \(\beta_{q} \in \{0.1, 0.4, 1.0, 2.0\}\) for our experiments.

\noindent In all of the above cases, the distribution parameter \(\beta\) used to generate diverse levels of data skew has the following property: a smaller \(\beta\) value results in a greater heterogeneity among the clients.

\subsubsection{Parameters and Grids}\label{sec:hyperparams}

We focus on server learning rate and momentum on the above experimental settings and perform both LHO and GHO (ground truth) grid search. Table~\ref{tab:grids} presents the grids used in all our experiments.

\begin{table}[!h]
\small
\caption{Hyperparameter (HP) grids for LHO and GHO.}
\centering
\label{tab:grids}
\begin{tabular}{cc}
\toprule
\multicolumn{1}{c}{\textbf{HP}} & \textbf{Grid}                       \\ \midrule
learning rate        & \begin{tabular}[c]{@{}l@{}}\textbf{iid} $\rightarrow$ {[}0.001, 0.01, 0.05, 0.1, 0.2, 0.3, 0.5{]}\\ \textbf{non-iid} $\rightarrow$ {[}0.01, 0.03, 0.05,  0.1, 0.3, 0.5{]}\end{tabular} \\ \midrule
momentum    & \begin{tabular}[c]{@{}l@{}} \textbf{iid} $\rightarrow$ {[}0, 0.3, 0.6, 0.9, 0.92, 0.93, 0.95{]}\\ \textbf{non-iid} $\rightarrow$ {[}0, 0.3, 0.6, 0.9{]}\end{tabular}           \\ \bottomrule
\end{tabular}
\end{table}

We keep the remaining parameters constant: We fix the client learning rate to $0.01$, the client momentum to $0.01$, the number of epochs to $e=30$ (for both LHO and GHO) and the number of local iterations to $\ell=2$ for GHO experiments. We keep $10\%$ of the training data as a validation set. For the aforementioned grid sizes of learning rate and momentum, each experimental setting for iid and non-iid requires $7*7=49$ and $6*4=24$ rounds of training, respectively. Consequently, the LHO experiments require $N*49$ and $N*24$ rounds of training for each client. Altogether, we performed over $4,067$ experiments for the iid and $6,912$ experiments for the non-iid experimental settings.

\subsection{Experimental Results}\label{sec:results}


\begin{figure*}[t]
\centering
    \begin{subfigure}[b]{.33\textwidth}
    \centering
    \includegraphics[width=\textwidth]{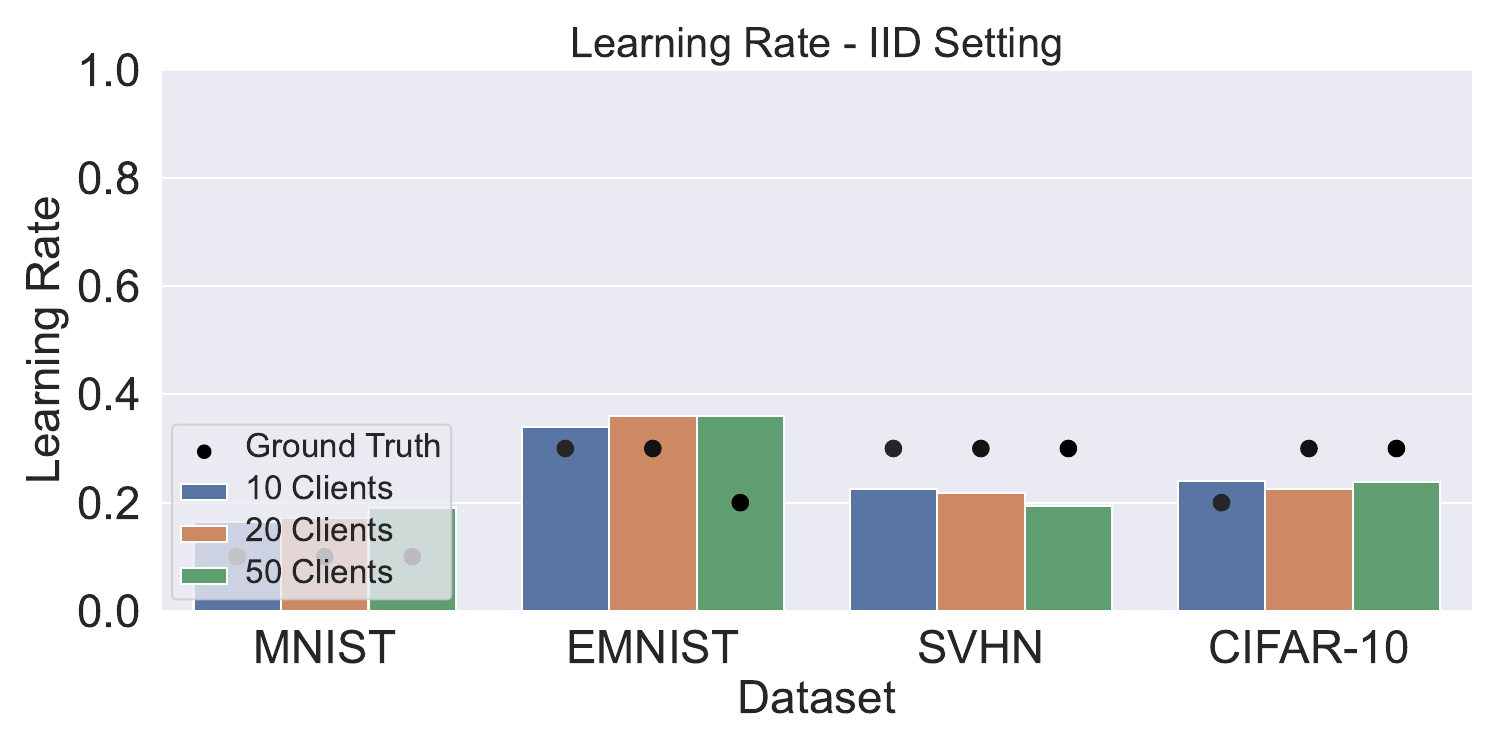} 
    \captionsetup{width=\textwidth}

    \caption{Learning Rate}
    \label{fig:lr_iid}
    \end{subfigure}
    \hfill
    \begin{subfigure}[b]{.33\textwidth}
        \centering
    \includegraphics[width=\textwidth]{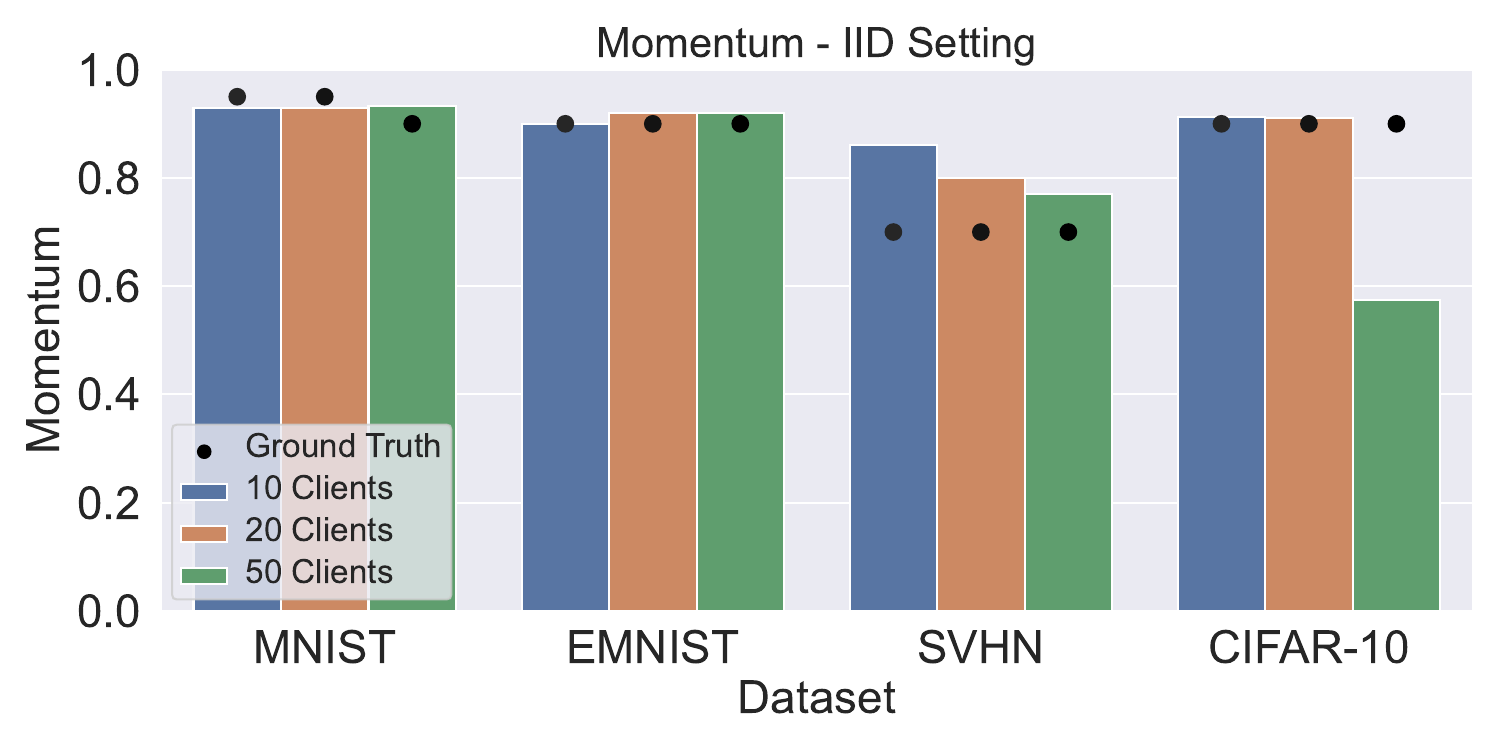}
    \captionsetup{width=\textwidth}

    \caption{Momentum}
    \label{fig:mom_iid}
    \end{subfigure}
    \hfill
    \begin{subfigure}[b]{.33\textwidth}
        \centering
    \includegraphics[width=\textwidth]{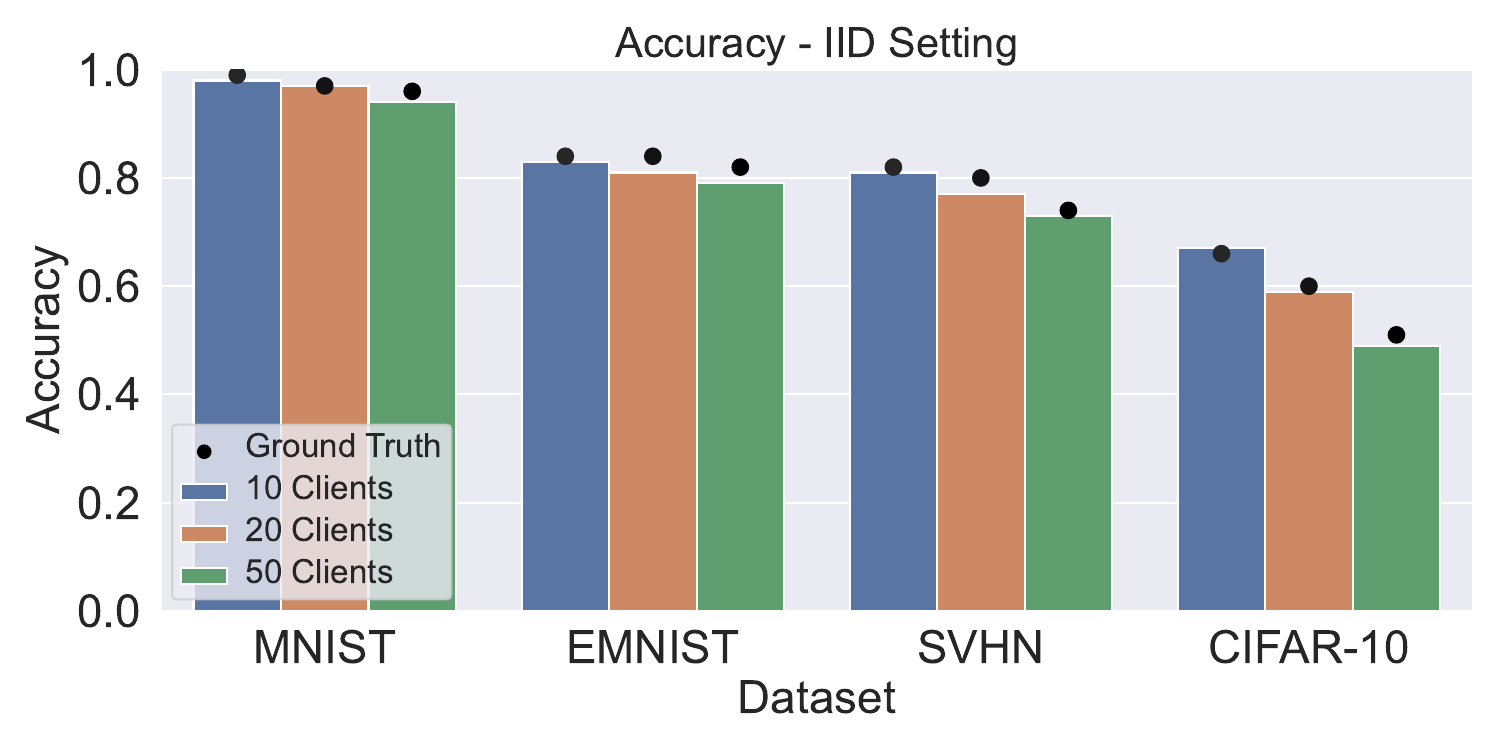}
    \captionsetup{width=\textwidth}

    \caption{Accuracy}
    \label{fig:acc_iid}
    \end{subfigure}

  \caption{Barplots of learning rate, momentum, and accuracy for the iid setting with \textbf{Mean} as the $\textsf{Combine}(\cdot)$ strategy, on various datasets and experiments. The ground truth (GHO result) is indicated by black dots. The remaining bar colors represent the results of combining the client optimal HPs with the \textbf{Mean} $\textsf{Combine}(\cdot)$ strategy, for variable number of clients ($N$).}
    \label{fig:iid}
\end{figure*}

\begin{figure*}[t]
\centering
    \begin{subfigure}[c]{\columnwidth}
    \centering
    \includegraphics[width=\textwidth]{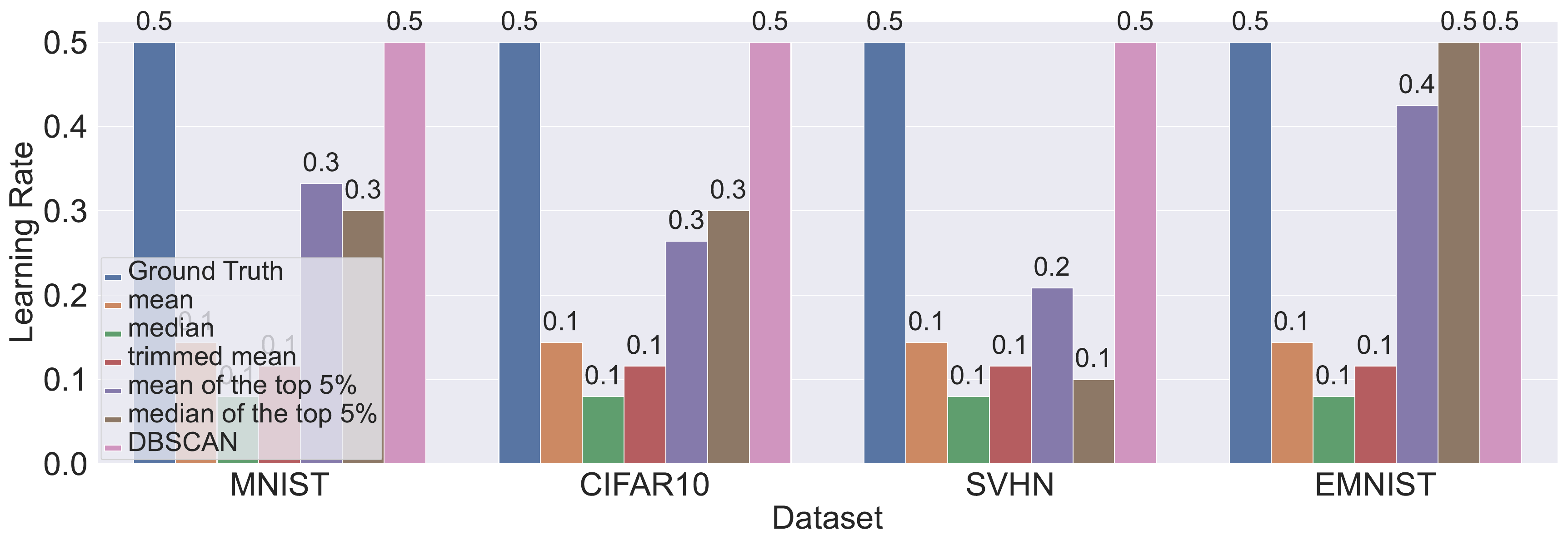} 
    \captionsetup{width=\textwidth}
     
    \caption{Learning rate with feature skew ($\beta_f=0.02$)}
    \label{fig:measurement_feature_lr}
    \end{subfigure}
    \begin{subfigure}[c]{\columnwidth}
    \centering
    \includegraphics[width=\textwidth]{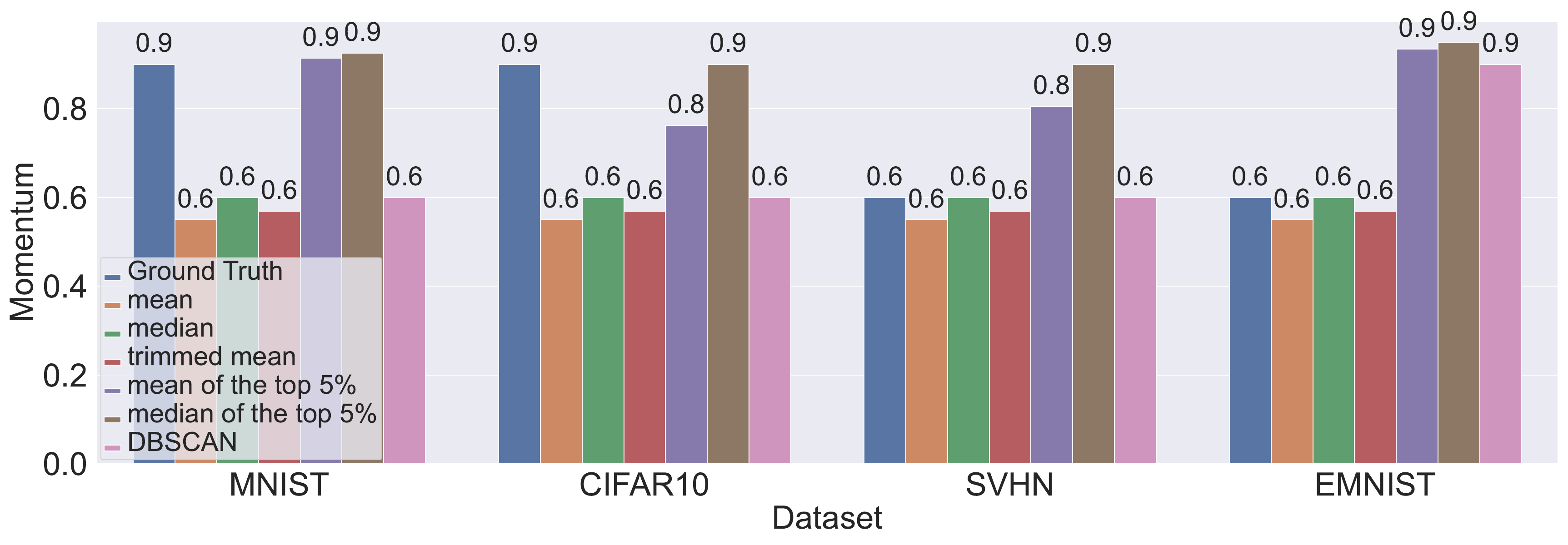}
    \captionsetup{width=\textwidth}
     
    \caption{Momentum with feature skew ($\beta_f=0.02$)}
    \label{fig:measurement_feature_mom}
    \end{subfigure}
    \begin{subfigure}[c]{\columnwidth}
    \centering
    \includegraphics[width=\textwidth]{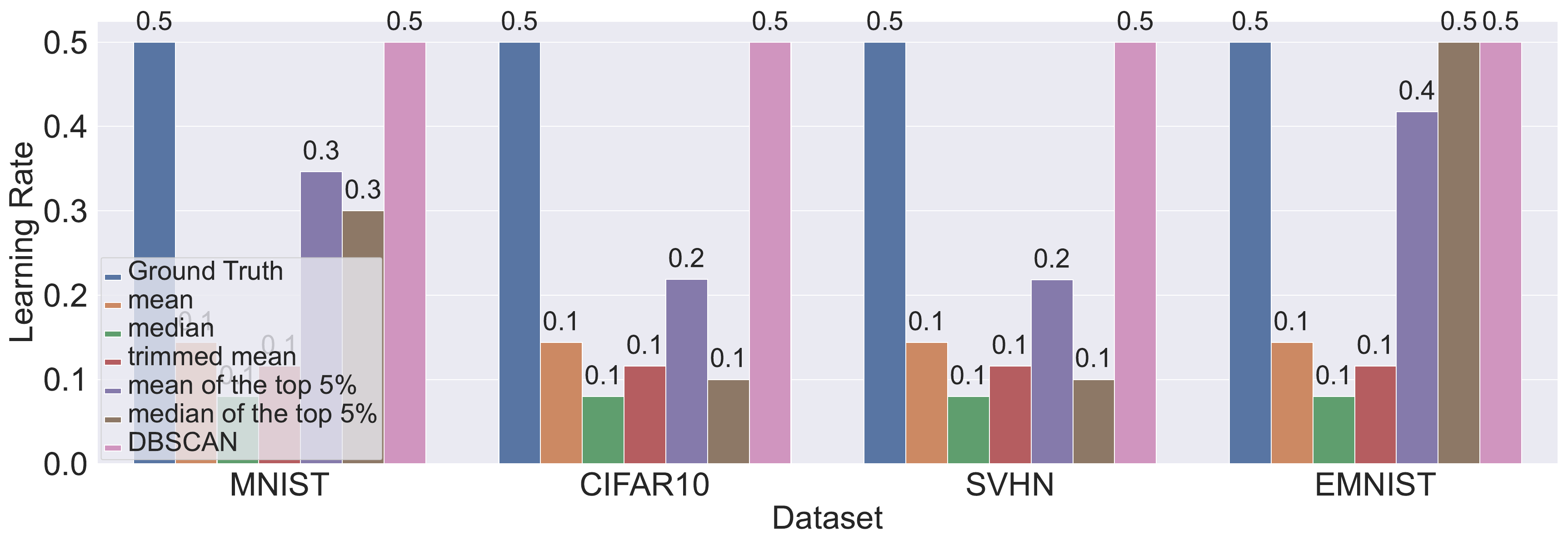} 
    \captionsetup{width=\textwidth}
     
    \caption{Learning rate with label skew ($\beta_{\ell}=1.0$)}
    \label{fig:measurement_label_lr}
    \end{subfigure}
    \begin{subfigure}[c]{\columnwidth}
    \centering
    \includegraphics[width=\textwidth]{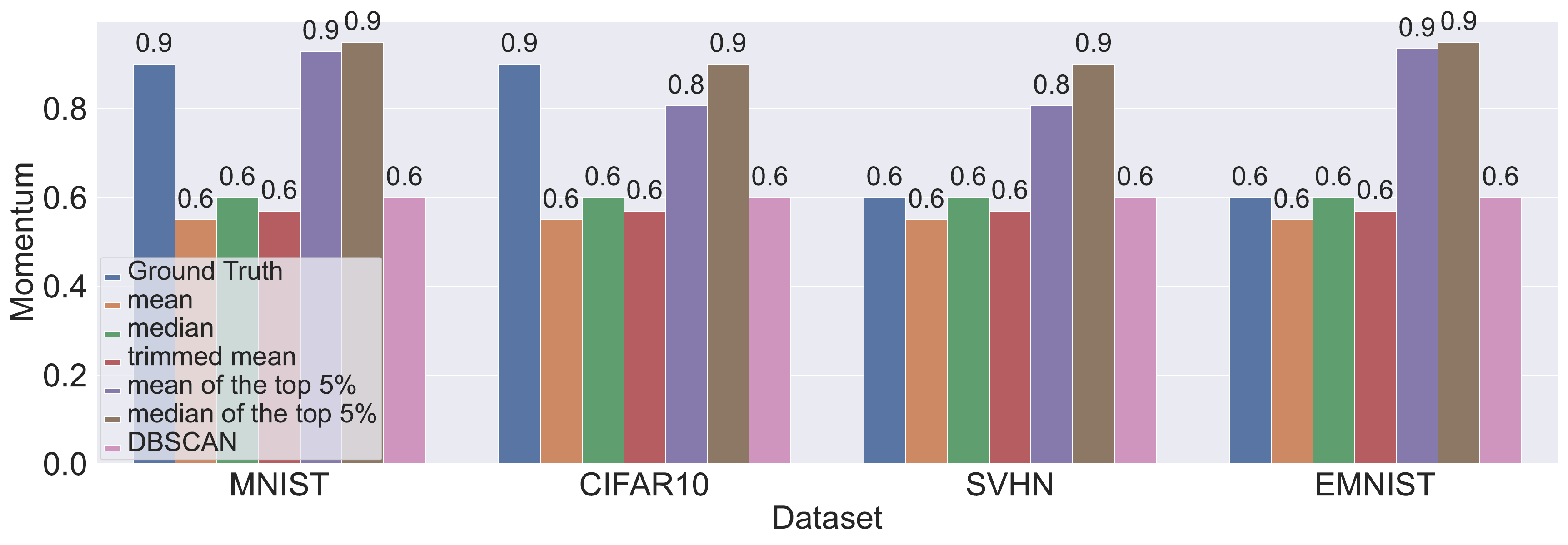} 
    \captionsetup{width=\textwidth}
     
    \caption{Momentum with label skew ($\beta_{\ell}=1.0$)}
    \label{fig:measurement_label_mom}
    \end{subfigure}
    \begin{subfigure}[c]{\columnwidth}
    \centering
    \includegraphics[width=\textwidth]{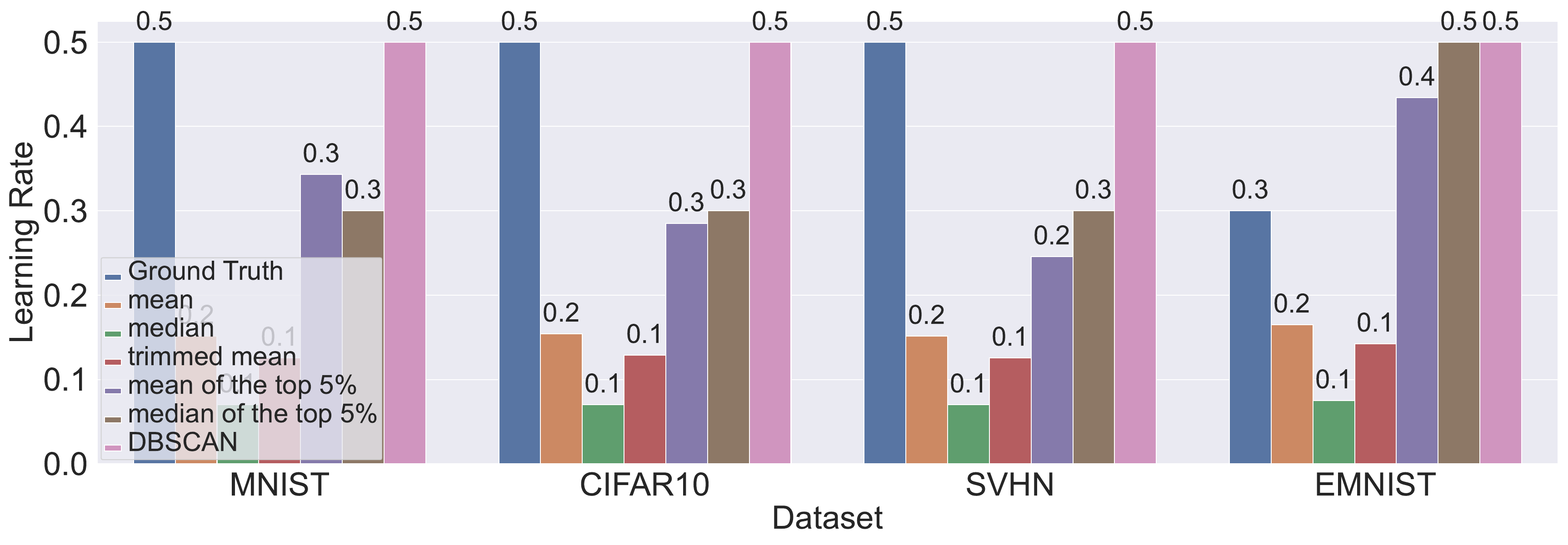} 
    \captionsetup{width=\textwidth}
     
    \caption{Learning rate with quantity skew ($\beta_q=0.4$)}
    \label{fig:measurement_qty_lr}
    \end{subfigure}
    \begin{subfigure}[c]{\columnwidth}
    \centering
    \includegraphics[width=\textwidth]{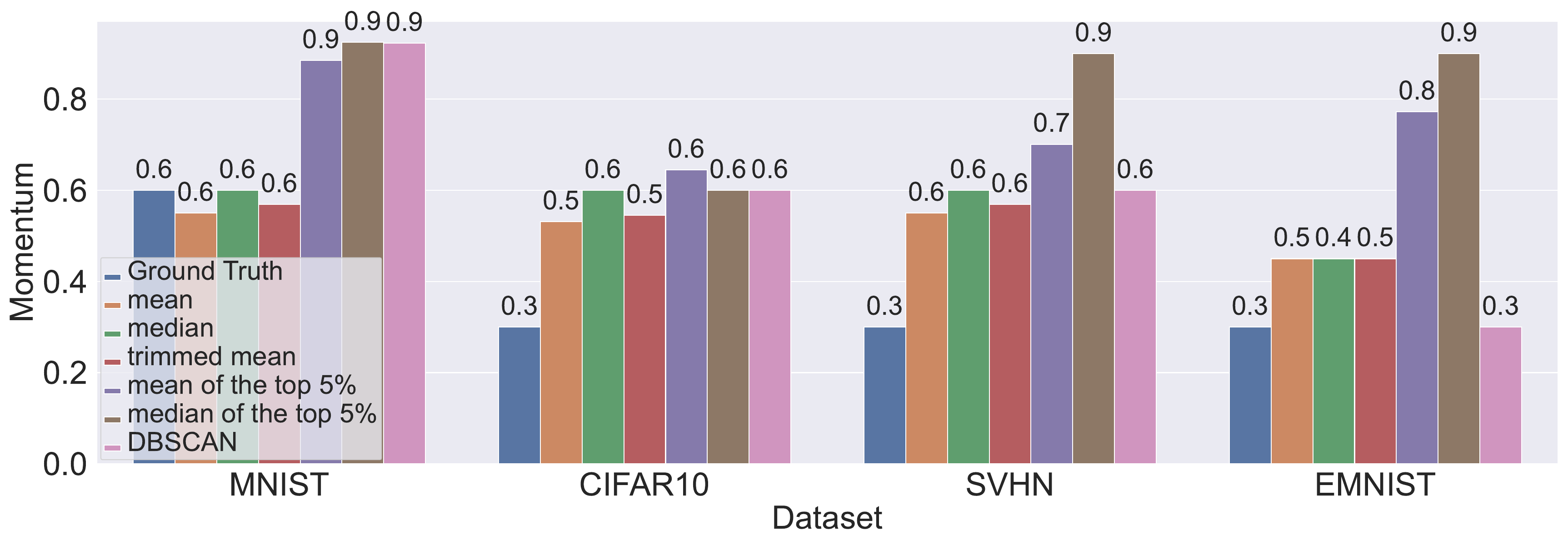}
    \captionsetup{width=\textwidth}
     
    \caption{Momentum with quantity skew ($\beta_q=0.4$)}
    \label{fig:measurement_qty_mom}
    \end{subfigure}
    
    \caption{Barplots of learning rate and momentum for the non-iid setting with $N=20$ clients, variable datasets and $\textsf{Combine}(\cdot)$ strategies. The top-row results are for feature skew ($\beta_f=0.02$), middle-row results for label skew ($\beta_{\ell}=1.0$), and bottom-row results are for quantity skew ($\beta_q=0.4$). Bar colors represent the HPs derived using various combination strategies.}
    \label{fig:measurement}
    
\end{figure*}

We showcase the results of our measurement study on both iid (Section~\ref{sec:results-iid}) and non-iid (Section~\ref{sec:results-noniid}) cross-silo federated learning settings. Then, we compare the tuning performance of the strategies identified by our measurement study with the state-of-the-art efficient HP tuning method for FL, FLoRA (Section~\ref{sec:comparison-flora}). Finally, we discuss the main takeaways of our measurement study (Section~\ref{sec:takeaways}).


\begin{figure*}[t]
\centering
    \begin{subfigure}[b]{.33\textwidth}
    \centering
    \includegraphics[width=\textwidth]{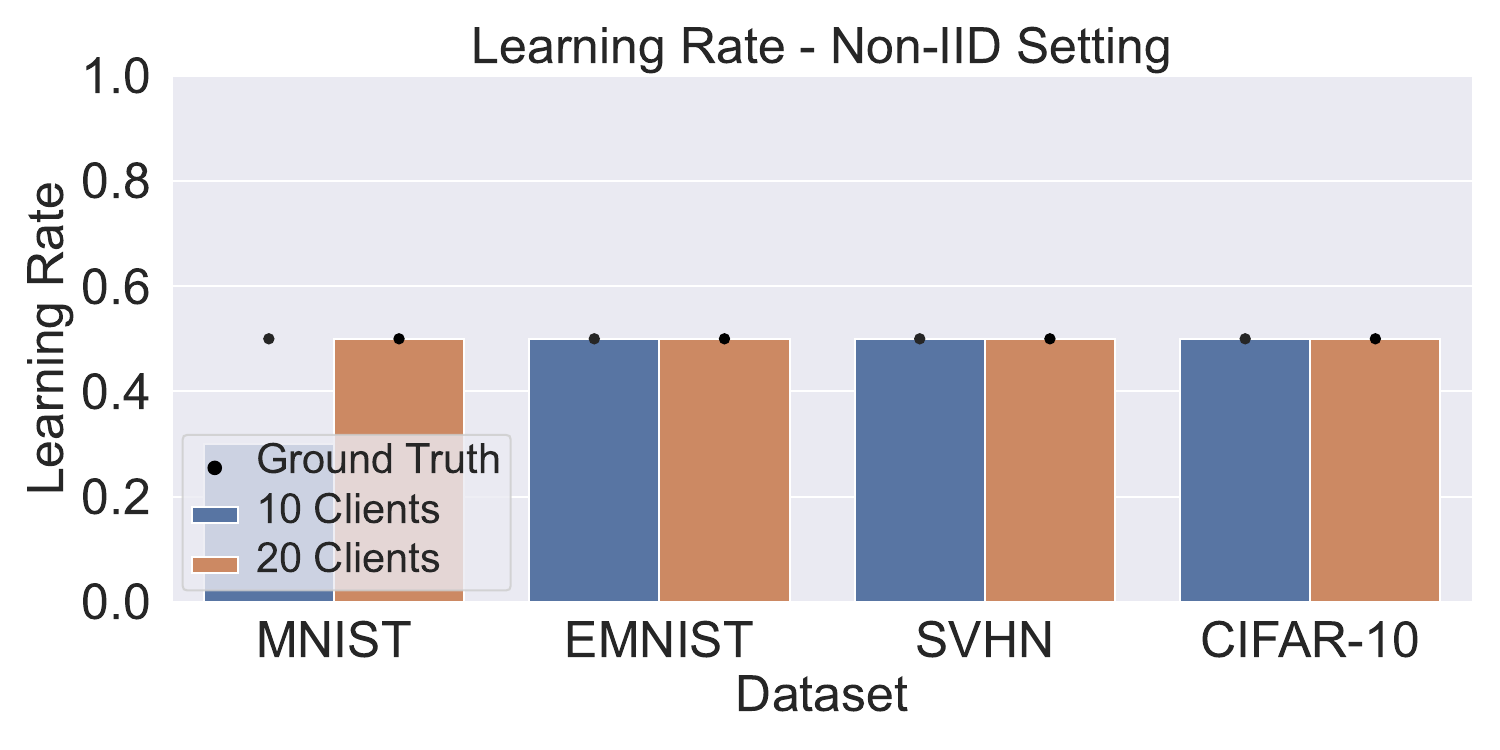}
    \captionsetup{width=\textwidth}
    \caption{Learning Rate}
    \end{subfigure}
    \hfill
    \begin{subfigure}[b]{.33\textwidth}
    \centering
    \includegraphics[width=\textwidth]{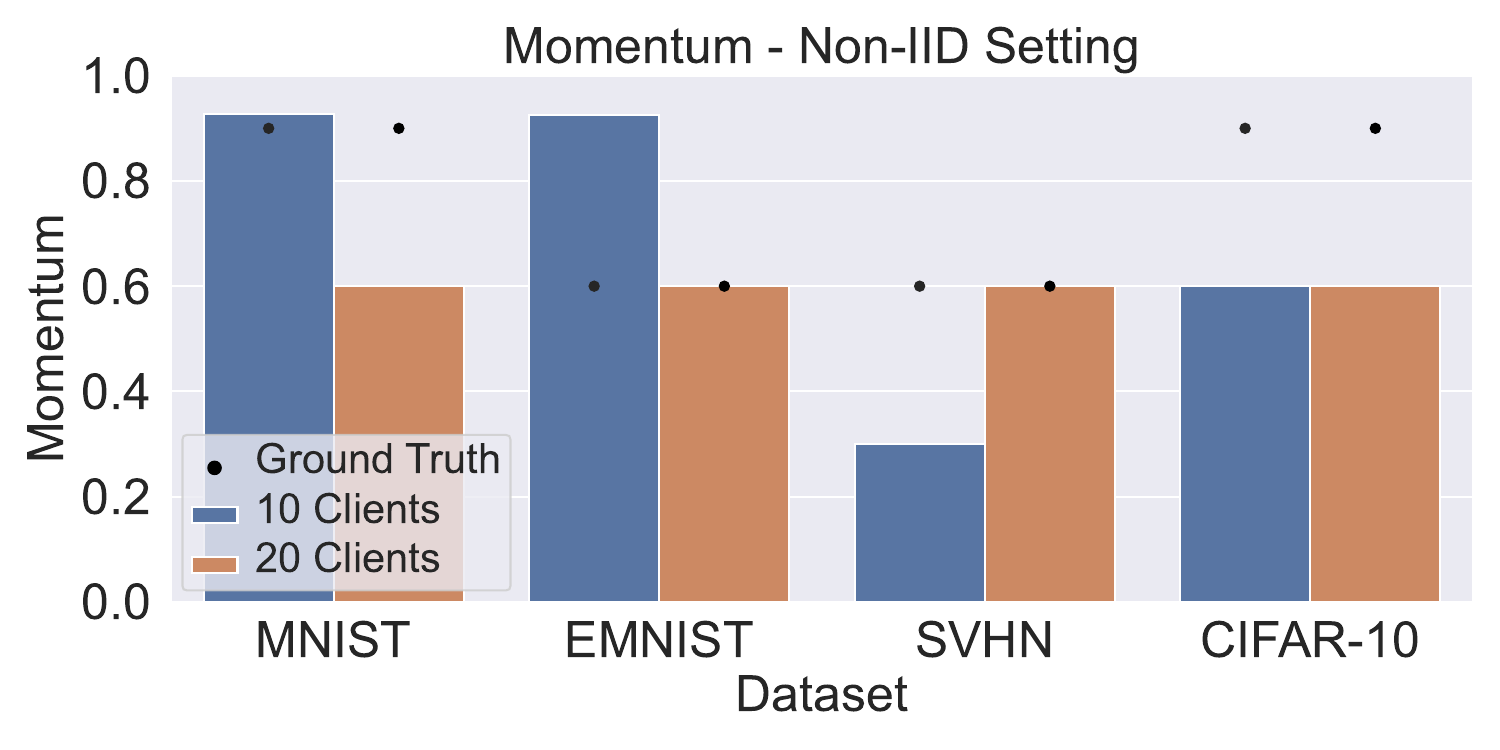}
    \captionsetup{width=\textwidth}
    \caption{Momentum}
    \end{subfigure}
    \hfill
    \begin{subfigure}[b]{.33\textwidth}
    \centering
    \includegraphics[width=\textwidth]{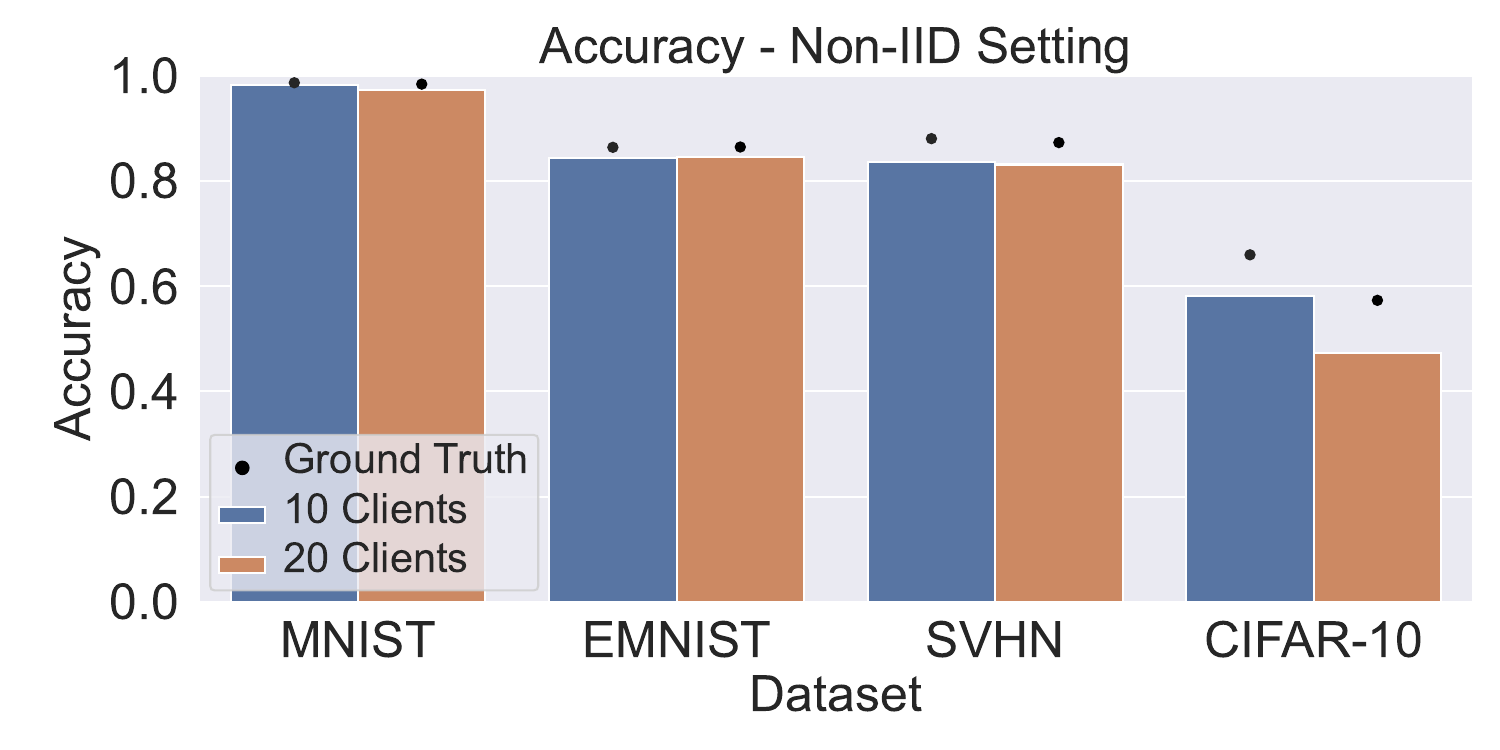} 
    \captionsetup{width=\textwidth}
    \caption{Accuracy}
    \label{fig:noniid-label-accuracy}
    \end{subfigure}
    \caption{Barplots of learning rate, momentum, and accuracy, for the non-iid setting with label skew ($\beta_{\ell}=1.0$), using DBSCAN as the $\textsf{Combine}(\cdot)$ strategy on various datasets and experiments. The ground truth (GHO results) is indicated by black dots. Bar colors represent the results of combining the client optimal HPs with the DBSCAN strategy, for variable number of clients ($N$).}
    \label{fig:noniid-label}
\end{figure*}

\begin{figure*}[t]
\centering
    \begin{subfigure}[b]{\columnwidth}
    \centering
    \includegraphics[width=\textwidth]{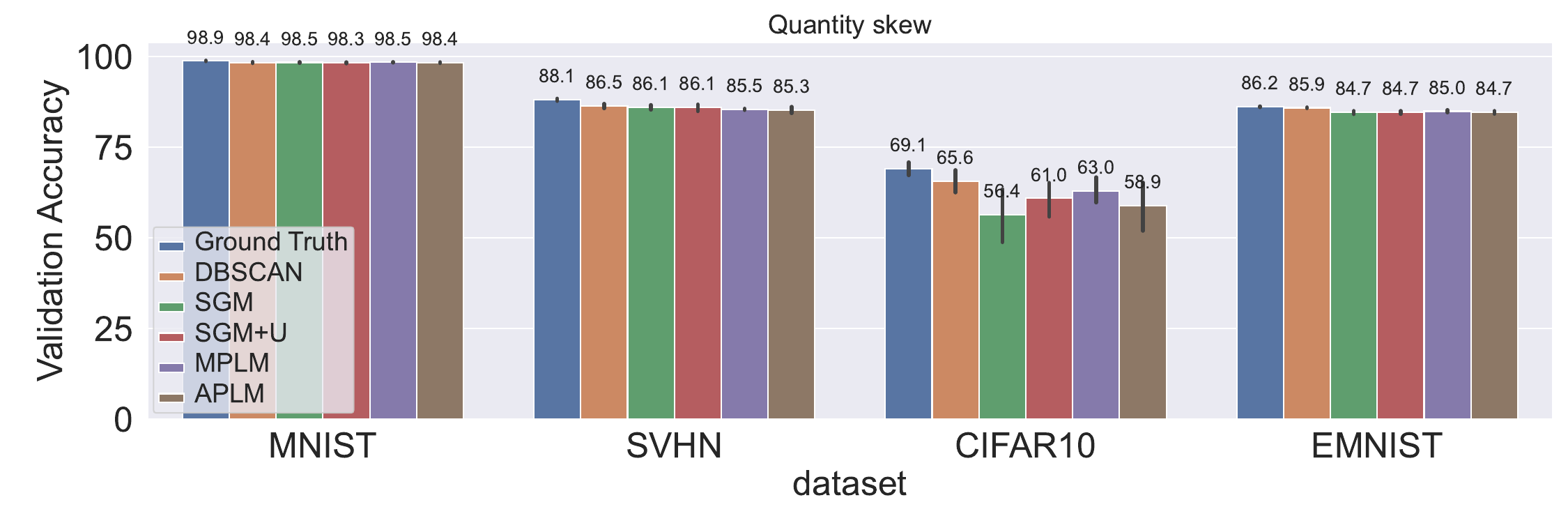} 
    \captionsetup{width=\textwidth}
     
    \caption{Validation accuracies -- averaged across the quantity skew setting.}
    \label{fig:qty}
    \end{subfigure}
    \hfill
    \begin{subfigure}[b]{\columnwidth}
    \centering
    \includegraphics[width=\textwidth]{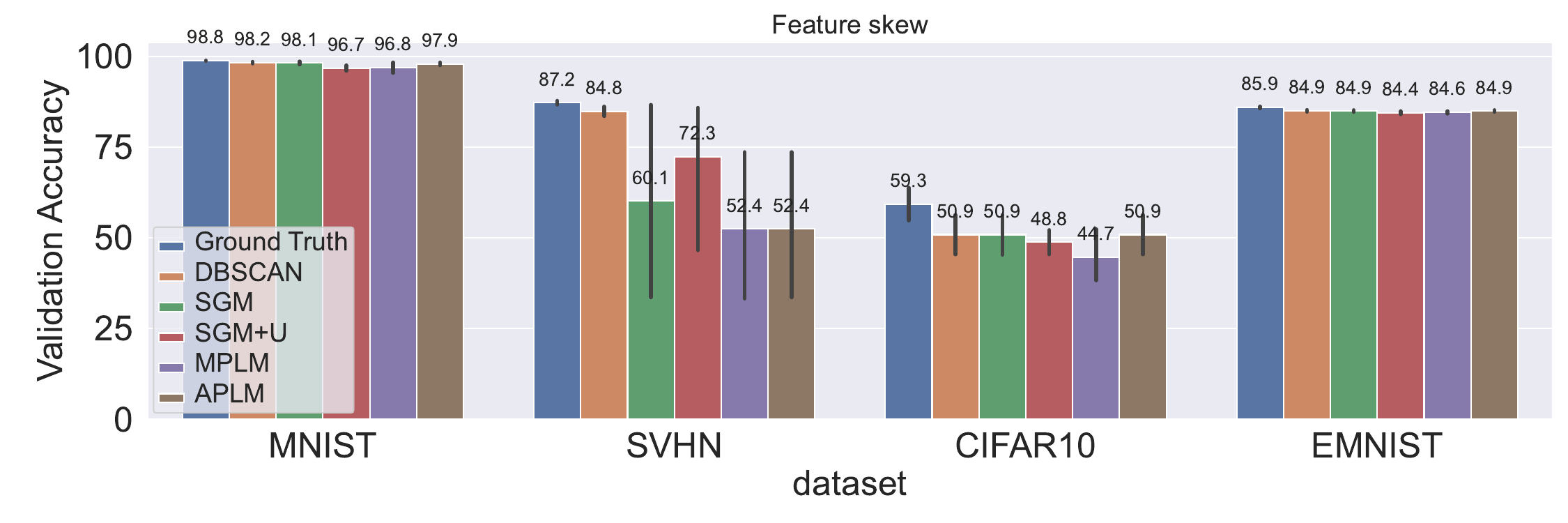}
    \captionsetup{width=\textwidth}
     
    \caption{Validation accuracies -- averaged across the feature skew setting.}
    \label{fig:ftr}
    \end{subfigure}
    \hfill
    \begin{subfigure}[b]{\columnwidth}
    \centering
    \includegraphics[width=\textwidth]{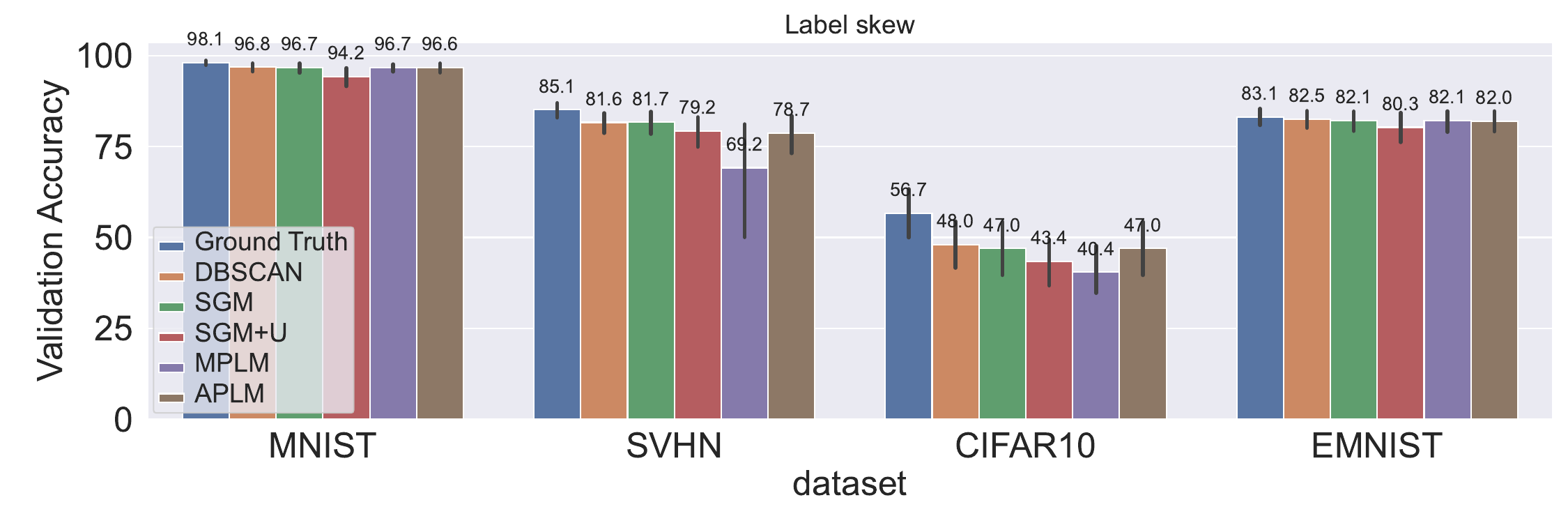} 
    \captionsetup{width=\textwidth}
     
    \caption{Validation accuracies -- averaged across the label skew setting.}
    \label{fig:lab}
    \end{subfigure}
    \hfill
    \begin{subfigure}[b]{\columnwidth}
    \centering
    \includegraphics[width=\textwidth]{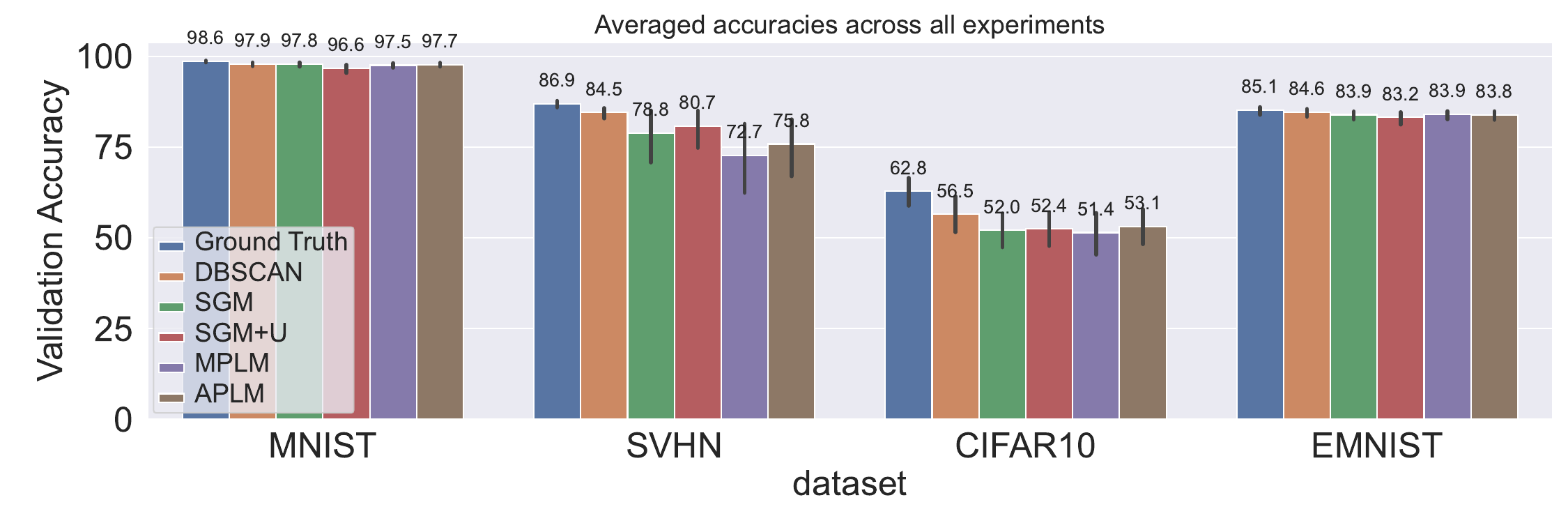}
    \captionsetup{width=\textwidth}
     
    \caption{Validation accuracies -- averaged across all skews.}
    \label{fig:acc}
    \end{subfigure}
    \caption{Comparison of the various combination strategies proposed in FLoRA~\cite{flora} vs. the density-based clustering strategy identified in our measurement study. The blue bar indicates the federated grid search results as ground truth. Each plot represents a different skew type (averaged across skew parameters with 10 and 20 clients).}
    \label{fig:noniid-comparison}
\end{figure*}

\subsubsection{IID Setting Results}\label{sec:results-iid}

Figure~\ref{fig:iid} displays our results on the iid setting for learning rate and momentum with $N=[10, 20, 50]$ clients using the \textbf{Mean} strategy as the $\textsf{Combine}(\cdot)$ method. We denote the ground truth (GHO results) with black dots. We observe that the learning rate or momentum determined by applying the \textbf{Mean} strategy on the client HPs closely approximates the ground truth, and the resulting HPs do not hinder the achieved test accuracy (Figure~\ref{fig:acc_iid}). We note that obtaining the optimal learning rate and momentum on the CIFAR-10 dataset is more challenging than on other datasets; we speculate that this is due to the more complex task of classifying RGB images. Moreover, we observe that discovering the optimal server HPs from the client HPs is more difficult with increasing number of clients; this is because the dataset size at each client decreases as the number of clients increase, i.e., the HPs that they locally discover are less \textit{optimal}. This is also reflected on Figure~\ref{fig:acc_iid}, which shows that the test accuracy slightly decreases with increasing number of clients, however, the HPs obtained by the \textbf{Mean} strategy yield similar accuracy to the (GHO) ground truth. Finally, we note that other combination strategies (e.g., median or density-based clustering) worked equally well for the iid setting indicating that in this case there is an inherent connection between the server HPs and the client ones. We omit the corresponding results due to space constraints.

\subsubsection{Non-IID Setting Results}\label{sec:results-noniid}

Figure~\ref{fig:measurement} showcases our measurement results for various $\textsf{Combine}(\cdot)$ strategies, datasets, and non-iid settings with $N=20$ clients: (i) feature skew with $\beta_f=0.02$ (top-row), (ii) label skew with $\beta_{\ell}=1.0$ (middle-row), and (iii) quantity skew with $\beta_q=0.4$ (bottom-row). We provide the same experiments with different skew parameters in Appendix~\ref{sec:additionalFigures}, Figure~\ref{fig:measurement-2}. Overall, we observe that straightforward combination strategies, such as computing the mean, its trimmed version, or the median, yield HPs that are not close to the ground truth; this demonstrates the challenge of performing HP tuning in non-iid settings. However, we observe that other combination strategies perform better: Density-based clustering optimally derives the ground truth server learning rate across all skews and datasets, while the mean/median of the top $5\%$ performs well for estimating the server momentum. For instance, the density-based clustering combination strategy perfectly matches the ground truth for learning rate (Figures~\ref{fig:measurement_feature_lr},~\ref{fig:measurement_label_lr} and~\ref{fig:measurement_qty_lr}) as well as the momentum in experiments with the SVHN and EMNIST datasets (Figures~\ref{fig:measurement_feature_mom},~\ref{fig:measurement_label_mom}, and~\ref{fig:measurement_qty_mom}). Similarly, the mean/median of the top $5\%$ HPs achieves the best results for estimating the momentum when there is feature or label skew on the MNIST and CIFAR10 datasets. When we consider the joint optimization of learning rate and momentum, across all our experimental settings with various skew types and parameters, datasets, and number of clients, we find that, on average, density-based clustering outperforms the other combination strategies. In particular, when we check the absolute distance of the parameters estimated by each combination function from the ground truth, we observe that out of 144 total experiments, DBSCAN achieves the lowest absolute distance from the ground truth in 95 of them. The second best approach was the mean/median of the top $5\%$ with 25 experiments achieving the lowest absolute distance.



Having identified density-based clustering as a promising combination strategy for non-iid settings, we perform further experiments to investigate how various factors affect its performance. Figure~\ref{fig:noniid-label} shows its results for a label skew setting with $\beta_{\ell}=1.0$ while similar plots for feature and quantity skew can be found in Appendix~\ref{sec:additionalFigures} (Figures~\ref{fig:noniid-feature} and~\ref{fig:noniid-quantity}). We observe that the learning rate or momentum determined through density-based clustering closely approximates the ground truth optimal HP with a slightly higher deviation for momentum values. This deviation could be due to the utilization of a smaller grid for momentum experiments, which results in more pronounced shifts from one value to another. Nonetheless, the resulting HP adjustments do not have a substantial effect on the achieved test accuracy, as illustrated in Figure~\ref{fig:noniid-label-accuracy} (and Figures~\ref{fig:noniid-feature-accuracy} and~\ref{fig:noniid-quantity-accuracy} for feature and quantity skew, resp.). For example, in the quantity skew setting, the test accuracy achieved with the parameters determined by density-based clustering is exactly the same as that achieved with the ground truth parameters (GHO) in 7 out of 8 experiments (Figure~\ref{fig:noniid-quantity-accuracy}). Moreover, we observe that the number of clients does not significantly affect the optimal HPs. This might be due to the fact that the number of samples in each dataset is sufficient to scale to 20 clients without a substantial accuracy loss in vanilla FL and this trend is observable in our settings.

\subsubsection{Comparison to Prior Work}\label{sec:comparison-flora}

We now perform a comparison between the best performing combination strategy, i.e., density-based clustering, and the closest related work, FLoRa~\cite{flora}. FLoRA~\cite{flora} is an HP tuning solution for federated learning that leverages local HP tuning to derive optimal global HPs (\(\theta ^{*}\)). Given its one-shot nature, and since it has been successfuly applied to non-iid settings, it is the most pertinent solution for a comparative analysis. 

In FLoRA, each client \(i\) performs local HPO and sends (HPs, validation loss) pairs \(E_i\) to the aggregator. Then, the aggregator constructs a unified loss surface \(l: \Theta \rightarrow \mathbb{R}\) using the pairs \(E_i\) and finds the best HP \(\theta ^{*} \leftarrow argmin_{\theta \in \Theta}l(\theta)\). The authors propose four ways of constructing such loss surfaces, which we re-implement within our experiments (LHO results) for a fair comparison:
\begin{itemize}[leftmargin=*]
    \item \textbf{Single global model (SGM).} All \(E_i\) sets are merged and used to train a global regressor \(f:\Theta \rightarrow \mathbb{R}\). The loss surface is defined as \(l(\theta) = f(\theta) \).
    \item \textbf{Single global model with uncertainty (SGM+U).} All \(E_i\) sets are merged and used to train a global regressor that provides uncertainty quantification around its predictions \(f:\Theta \rightarrow \mathbb{R}\), \(u:\Theta \rightarrow \mathbb{R}_{+}\), where \(f(\theta)\) is the mean prediction of the model at \(\theta\) and \(u(\theta)\) quantifies the uncertainty around this prediction. The loss surface is defined as \(l(\theta) = f(\theta) + \alpha * u(\theta)\) for some \(\alpha > 0\).
    \item \textbf{Maximum of per-party local models (MPLM).} This approach trains a regressor \(f_i:\Theta \rightarrow \mathbb{R}\) for each client set \(E_i\). The loss surface is defined as \(l(\theta) = max f_i(\theta)\).
    \item \textbf{Average of per-party local models (APLM).} This approach trains a regressor \(f_i:\Theta \rightarrow \mathbb{R}\) for each client set \(E_i\). The loss surface is defined as \(l(\theta) = average f_i(\theta)\).
\end{itemize}
\medskip

All trained regressors are Random Forest ones, except for SGM+U, which trains a Gaussian Process regressor with a radial basis function kernel. We compare the validation accuracies of the models trained with: (a) our derived global HPs, (b) FLoRA-derived global HPs, and (c) the optimal, global HPs found by GHO. Figure~\ref{fig:noniid-comparison} shows the average validation accuracies achieved among various skew types (Figures~\ref{fig:qty}, ~\ref{fig:ftr}, and~\ref{fig:lab}) and among all skew types (Figure~\ref{fig:acc}). Comparing the validation accuracy achieved with the HPs estimated by the density-based clustering combination strategy with that of the various FLoRA approaches, we deduce that on average, as well as for each skew individually, our strategy outperforms FLoRA for every dataset. The feature skew experiments for the SVHN dataset deem to be especially challenging for FLoRA (note the high uncertainty indicated by the error bars), with an average drop of $25\%$ in validation accuracy compared to our strategy. When compared to the optimal HPs found by GHO, our strategy achieves close to perfect results in all datasets except for CIFAR-10 which is the most challenging dataset in all experiments (see Section~\ref{sec:results-iid}). Moreover, we observe that the label skew (Figure~\ref{fig:lab}) is more challenging than other skews (Figures~\ref{fig:qty} and~\ref{fig:ftr}) which is compliant with the findings of Li et al.~\cite{li2021federated}.

\subsubsection{Takeaways}\label{sec:takeaways}

We conducted extensive experiments in both iid and non-iid cross-silo FL settings, aiming to identify a suitable $\textsf{Combine}(\cdot)$ strategy for deriving the server HPs from the client ones. We found that server HPs in iid settings exhibit a straightforward connection with the client ones and that a simple averaging combination strategy yields good enough performance (Figure~\ref{fig:iid}). In contrast, the non-iid setting poses a more challenging scenario for HP tuning, and simple combination strategies that estimate the mean, its trimmed version, or the median, do not derive optimal sets of server HPs. However, our experiments revealed that, on average, density-based clustering outperforms other combination strategies, while the mean/median of the top 5\% of client HPs uncovers the optimal momentum values. When experimenting with various types of non-iid settings, we found that performing HP tuning under label skew is the most challenging case (as reflected by the decrease in model accuracy compared to other skews). On the other hand, model accuracy under quantity skew exhibit robustness to parameter variations. This is because the data distribution remains consistent across clients (and only the quantity of samples varies between them). Regarding the datasets we employed for our study, we found that performing HP tuning on the EMNIST and CIFAR-10 datasets is more challenging than the rest. This is due to the fact that the classification tasks are harder given the variety of samples and outputs. Moreover, we observed that as long as the skew parameters are the same for the experiments, the number of clients in cross-silo FL settings does not significantly affect the optimal HPs. Finally, we compared the density-based clustering strategy uncovered by our measurement study with the state-of-the-art, FLoRA~\cite{flora}, on various non-iid settings, and we found that, on average, it is a much more effective strategy yielding a validation accuracy close to the model trained with the optimal HPs.

\section{Towards Privacy-Preserving Hyperparameter Tuning for Federated Learning}\label{sec:privFL}

Our benchmarking (Section~\ref{sec:design}) provided interesting insights into the connection between the client and server HPs in various federated learning settings. Moreover, it allowed us to establish simple and efficient techniques for tuning the server HPs by combining the optimal client HPs on the server side. However, revealing the optimal client HPs to the FL server opens up a new privacy attack surface, as prior research has shown that data samples with certain characteristics, e.g., outliers, can significantly influence the HP optimization process, and are hence vulnerable to inference attacks~\cite{papernot2022hyperparameter}. To address this issue, we introduce a novel framework called \sys that enables the development of various HP tuning techniques in a privacy-preserving manner by employing multiparty homomorphic encryption (MHE). We use our framework to implement and experimentally evaluate the two main strategies, i.e., averaging and density-based clustering, that our measurement study identified as suitable for iid and non-iid settings, respectively. Our framework is flexible and easily extensible to other HP tuning techniques that might be discovered by researchers. The rest of the section is organized as follows: We first provide background information on MHE (Section~\ref{sec:mhe}) and then we describe \sys and how it supports the execution of combination strategies for HP tuning among multiple clients (Section~\ref{sec:privtuna}). Finally, we experimentally evaluate \sys in terms of runtime and communication overhead, as well as HP tuning precision (Section~\ref{sec:experiments-privtuna}).



\subsection{Background: Multiparty Homomorphic Encryption}\label{sec:mhe}

Homomorphic encryption supports arithmetic operations to be carried out on encrypted data. Our framework relies on multiparty homomorphic encryption (MHE), a variant of homomorphic encryption (HE) for multi-party settings, to support the HP combination on the server side and to facilitate distributed cryptographic operations among the clients. Specifically, we utilize the multiparty version of Cheon-Kim-Kim-Song (CKKS), a leveled HE scheme based on the problem of ring learning with errors (RLWE) as introduced by Cheon et al.~\cite{cheon2017homomorphic} and extended to the multiparty setting by Mouchet et al.~\cite{mouchet2019distributedbfv}. This particular scheme offers plausible security against post-quantum attacks, supports floating-point arithmetic, and can withstand collisions of up to N-1 parties under an honest-but-curious threat model (anytrust model).

For a cyclotomic polynomial ring with a dimension $\mathcal{N}$ (a power-of-two integer), the MHE scheme establishes the plaintext and ciphertext space, denoted as $R_{Q}=\mathbb{Z}_{Q}[X]/(X^\mathcal{N}+1)$. Here, $Q$ serves as the ciphertext modulus at an initial level which represents the depth of homomorphic operations that can be performed on a ciphertext. A plaintext/ciphertext encodes a vector of values, up to a maximum of $\mathcal{N}/2$ to its slots. As such, the scheme enables parallel operations through single instruction, multiple data to $\mathcal{N}/2$ values encoded in a plaintext/ciphertext. We introduce below the primary operations used in our framework. Note that CKKS only allows for limited arithmetic operations, i.e., additions and multiplications. For other operations such as division or comparison, we rely on approximations.

\begin{itemize}[leftmargin=*]
    
    \item $\textsf{SecKeyGen}(1^\lambda)$: Returns the set of secret keys $\{sk_i\}$, i.e., $sk_i$ for each party $P_i$, given a security parameter $\lambda$.

   \item $\textsf{DKeyGen}(\{sk_i\})$: Returns the collective public key $pk$ and the evaluation keys $[ek]$.
    

    \item $\textsf{DDecrypt}(\bm{c}, \{sk_i\})$: Returns the plaintext $p$, the decryption of a ciphertext $\bm{c}$.
    
    
   \item $\textsf{Encrypt}(pk, {p})$: Returns $\bm{c}_{pk}\in R^2_{Q}$ such that the decryption results in $\textsf{DDecrypt}(\bm{c}_{pk}, \{sk_i\}) \approx {p}$.
     
   \item $\textsf{Add}(\bm{c}_{pk}, \bm{c'}_{pk})$: Returns $(\bm{c} + \bm{c'})_{pk}$.

   \item $\textsf{Divide}(\bm{c}_{pk},\bm{c'}_{pk})$: Returns $\bm{c}_{pk}/\bm{c'}_{pk}$ following Goldschmidt's iterative division algorithm~\cite{Goldschmidt}.
  
   \item $\textsf{Compare}(\bm{c}_{pk},\bm{t}_{pk})$: Compares $\bm{c}_{pk}$ and $\bm{t}_{pk}$ by using Cheon et al.'s comparison approximation~\cite{Cheon2020}. It returns a value in $[-1,1]$ by using several consecutive polynomial evaluations of two different polynomials $f$ and $g$ over the difference of two ciphertexts. If $\bm{c}_{pk}=\bm{t}_{pk}$, it returns 0.
   
   \item $\textsf{Sub}(\bm{c}_{pk}, \bm{c'}_{pk})$: Returns $(\bm{c} - \bm{c'})_{pk}$.
     
  \item $\textsf{Mul}_{\textsf{pt}}$($\bm{c}_{pk}, {p})$: Returns $(\bm{c} \cdot p)_{pk}$.
     
   \item $\textsf{Mul}_{\textsf{ct}}$($\bm{c}_{pk}, \bm{c'}_{pk})$: Returns $(\bm{c} \cdot \bm{c'})_{pk}$.
     
    
   \item $\textsf{DBootstrap}(\bm{c}_{pk}, \{sk_i\})$: Returns $\bm{c}_{pk}$ with initial level $L$ (i.e., it refreshes the ciphertext).
   
\end{itemize}

\noindent The functions above that start with `D’ are distributed, and executed among all the clients, while the others can be executed locally by any client (that holds the collective public key).

\subsection{\sys Description} \label{sec:privtuna}

We first provide a high-level overview of \sys (Section~\ref{sec:overview}). Then, we show how \sys enables the privacy-preserving execution of the two HP combination strategies, i.e., federated mean (Sections~\ref{sec:fedAvg}) and DBSCAN (Section~\ref{sec:fedDBSCAN}), that our measurement study identified as promising for iid and non-iid cross-silo federated learning settings, respectively. Then, we discuss how \sys can be extended to other HP tuning strategies (Section~\ref{sec:extensions}).

\subsubsection{\sys Overview}\label{sec:overview}

Recall that our approach for efficient HP tuning in cross-silo federated learning (Algorithm~\ref{algo:method}) relies on: (i) the clients to perform local hyperparameter optimization (LHO) on their dataset, and to send the resulting HP sets to the server, and (ii) the server to combine ($\textsf{Combine}(\cdot)$, Line 5, Algorithm~\ref{algo:method}) the received HP sets and derive the optimal global parameters for the federated training process. To enable this workflow in a privacy-preserving manner, \sys encrypts all LHO results, and executes the combination strategy \textbf{under encryption} leveraging on the MHE functionalities summarized in Section~\ref{sec:mhe}. As such, \sys ensures that neither the server nor other clients can directly access the HP values discovered by LHO at each client.

In more detail, \sys operates as follows: (i) The clients generate their secret keys ($\textsf{SecKeyGen}(1^\lambda)$) and interact with each other to generate a collective public key ($\textsf{DKeyGen}(\{sk_i\})$) and possibly additional evaluation keys required to execute the combination function under encryption. (ii) Each client encrypts its LHO results (i.e., the (hyperparameter, accuracy) pairs) with the collective public key and communicates the resulting ciphertexts to the server. (iii) The server executes the $\textsf{Combine}(\cdot)$ function on the received ciphertexts, leveraging on the homomorphic properties of the underlying encryption scheme. (iv) The server interacts with the clients that collectively decrypt ($\textsf{DDecrypt}(\bm{c}, \{sk_i\})$) the global HPs. We here note that evaluating certain $\textsf{Combine}(\cdot)$ strategies under encryption introduces challenges due to the limited arithmetic operations supported by the MHE scheme. In the following, we describe how to construct private versions of the federated mean and density-based clustering combination strategies in \sys.

\subsubsection{Private Federated Mean (PF-Mean)}\label{sec:fedAvg}

Computing the mean of the HPs discovered by LHO at each client requires their summation and a division by the number of contributed HP values. If the number of contributed HP values is known in advance (e.g., if each client only sends a single best HP set), then computing the mean under encryption is straightforward: It only requires homomorphic additions (i.e., $\textsf{Add}(\bm{c}_{pk}, \bm{c'}_{pk})$) and a homomorphic multiplication by a constant, i.e., the (plaintext) inverse of the number of clients, using $\textsf{Mul}_{\textsf{pt}}$($\bm{c}_{pk}, {p})$. However, in \sys, we assume that each client can contribute as many LHO results as they desire (e.g., their top $x\%$ HP sets based on validation accuracy). Thus, \sys executes the PF-Mean computation by letting each client encrypt the sum of each HP (over their HP sets) in the slots of one ciphertext, plus an additional ciphertext indicating how many HP sets they are contributing. Then, the server can compute the mean of each HP by performing additions that are inherently supported by the CKKS scheme and on the division operation ($\textsf{Divide}(\bm{c}_{pk},\bm{c'}_{pk})$)) that is possible through Goldschmidt's algorithm~\cite{Goldschmidt}. Note that excluding the key generation protocol ($\textsf{DKeyGen}(\{sk_i\})$), the PF-Mean computation in \sys requires a single communication round: Each client sends its optimal HPs encrypted to the server, the server performs the summation and the division, and returns the encrypted result back to the clients for collective decryption.

\subsubsection{Private Federated DBSCAN (PF-DBSCAN)}\label{sec:fedDBSCAN}

Performing density-based clustering under encryption at the server side (similar to the PF-Mean strategy of Section~\ref{sec:fedAvg}) is a very challenging task; DBSCAN requires the (homomorphic) evaluation of multiple non-linear operations, i.e., comparisons, which makes its privacy-preserving realization impractical. To avoid this issue, we leverage on a federated density-based clustering algorithm~\cite{marino2022} that enables the clustering to be performed collectively by all the clients of the federation. This algorithm involves partitioning the feature (i.e., hyperparameter) space into a grid with cells of a fixed granularity \(L\), a parameter analogous to $\mathcal{E}$ in traditional DBSCAN (see Appendix~\ref{subsec:dbscan}). Rather than sharing its raw HPs, each client communicates the number of points within the grid cells to the aggregation server. The server aggregates this information by summing up the number of points in each cell and uses the $\textsf{MinPts}$ parameter to create and expand clusters across neighboring cells. Finally, each client receives the cluster label of each cell within the feature space. We refer the reader to~\cite{marino2022} for more details. In the following, we enhance this algorithm with privacy protection using MHE and we modify its workflow by offloading operations that are not natively supported by the CKKS scheme to the client side.



Algorithm~\ref{algo:feddbscan} displays the privacy-enhanced version of the federated DBSCAN. The process starts after the clients have performed LHO to obtain their respective $(\{hp\}_i,\{accuracy\}_i)$ pairs and after they have established their cryptographic keys (secret keys and the collective public key -- see Section~\ref{sec:overview}). The algorithm proceeds in four stages:

\descr{Grid Setup.} Initially, each client performs a pre-processing step to retain the HPs with the best validation accuracy, which it scales to $[0,1]$ using a MinMax scaler. This step is pivotal for a homomorphic comparison operation required later in the algorithm. Then, each client performs discretization (Line 3, Algorithm~\ref{algo:feddbscan}) and creates a grid of cells represented as a matrix $M_i$; this matrix is populated by the total number of points inside each cell. A point is determined by the HP values, e.g., learning rate and momentum define a 2-dimensional point. Subsequently, each client computes \textit{the closest cell map} $Cell_i$ for each point to retain a mapping \textit{point --> closest cell}, where the closest cell is defined as the shortest Euclidean distance between the point and its neighboring (top, bottom, left, right) cells. To prevent the leakage of the HP grids, each client encrypts $M_i$ and sends it to the server (Lines 5 and 6, Algorithm~\ref{algo:feddbscan}). Note that by using the encoding (packing) capability of the MHE scheme, one matrix is encrypted into a single ciphertext, where the ciphertext slots represent the matrix entries.

\descr{Aggregation -- Round 1.} After the collection of the encrypted $M_i$ matrices from the clients, the server aggregates (i.e., homomorphically sums) the grids and homomorphically compares each matrix cell with the threshold $\textsf{MinPts}$ to discover the \textit{dense} cells (Line 9, Algorithm~\ref{algo:feddbscan}). To enable this operation under encryption, \sys employs the approximate comparison scheme proposed by Cheon et al.~\cite{Cheon2020}. Then, the server sends the (encrypted) dense mask (including 0s for sparse cells and 1s for dense cells) to the clients that collectively decrypt it (Line 11, Algorithm~\ref{algo:feddbscan}).

\descr{Local Clustering.} Upon receiving the dense cell mask $D$, each client identifies its local dense cells by multiplying it with $M_i$. If no data points fall within a dense cell, clients refer to $Cell_i$ to validate whether the mapped ``closest cell" qualifies as dense. If it does, the client retains the data point, assigning its value to the nearest cell; otherwise, the point is excluded. This step filters data points eligible for cluster formation, i.e., those within dense cells and those mapping to dense cells via the \textit{closest cell map}. Note that for the non-dense points that are retained, their values are assigned to the nearest adjacent dense cell; this measure is taken to prevent a scenario in which two clusters become interconnected through a data point originating from a non-dense cell. Then, each client merges adjacent cells by iteratively selecting a cell, appending all its neighbors to a queue, and checking if the cell in the queue is zero/non-zero (Line 15, Algorithm~\ref{algo:feddbscan}). Subsequently, each client fills the dense cells with the HPs and accuracy values associated with the retained data points. As such, each client creates a matrix for each HP and an additional one for the corresponding validation accuracies (Line 16, Algorithm~\ref{algo:feddbscan}). Finally, the clients record the total number of points in each cluster (required for averaging and deriving the final HP values). Each client encrypts their HPs and accuracies into a single ciphertext leveraging the packing capability of CKKS and encrypts the final count matrices. These two ciphertexts are sent to the server (Lines 18-20, Algorithm~\ref{algo:feddbscan}).

\descr{Aggregation -- Round 2.} The server aggregates the received matrices and computes the average HPs and the corresponding accuracy to find the final global configuration on the single packed ciphertext (Lines 22-25, Algorithm~\ref{algo:feddbscan}). As the averaging requires division, we employ the $\textsf{Divide}$ operation using Goldschmidt's iterative algorithm~\cite{Goldschmidt}. The averaged HP and accuracies are transmitted to clients, who then collaboratively decrypt this result. Finally, the clients use the decrypted average accuracy matrix to rank and obtain the final HP values (Line 28, Algorithm~\ref{algo:feddbscan}).

\begin{algorithm}[t]
\caption{Private Federated DBSCAN (PF-DBSCAN) Algorithm.}
\label{algo:feddbscan}
\begin{algorithmic}[1]
\Statex \textbf{Grid Setup:}
\For{Each client $S_i$ with $i \in \{1,...,N\}$}
\State Preprocessing: MinMax scaling of HPs
\State $M_i \gets $ \textsf{Discretize}($\{hp\}_i)$:
\State Precompute \textit{the closest cell map} $Cell_i$
\State $M_i \gets \textsf{Encrypt}(pk, M_i)$
\State Send $M_i$ to the server
\EndFor
\Statex \textbf{Aggregation -- Round 1:} \Comment{Server executes}
\State ${M} \gets \textsf{Add}(M, M_i)$, $\forall i \in \{1, \dots, N\}$
\State Create dense cell mask: $D \gets \textsf{Compare}(M, \textsf{MinPts})$
\State Send $D$ to the clients
\State $D \gets \textsf{DDecrypt}(D, \{sk_i\})$
\Statex \textbf{Local Clustering:}
\For{Each client $S_i$ with $i \in \{1,...,N\}$} 
\State $LD_i \gets D \odot M_i$
\State Move non-dense \textit{points} to the closest dense cell
\State Merge adjacent cells to form final local clusters
\State Create grid $HP_{ij}$, $\forall j \in \{hp\}_i$, and accuracies $A_i$
\State Create matrix $T_i$ with total number of points in each cluster
\State $HP_{ij} \gets \textsf{Encrypt}(pk, HP_{ij})$, $\forall j \in \{hp\}_i$
\State $A_{i} \gets \textsf{Encrypt}(pk, A_{i})$, $T_{i} \gets \textsf{Encrypt}(pk, T_{i})$
\State Send $HP_{ij}, \forall j \in \{hp\}$, $A_{i}$, $T_{i}$ to the server
\EndFor
\Statex \textbf{Aggregation -- Round 2:} \Comment{Server executes}
\State ${HP_j} \gets \textsf{Add}(HP_{j}, HP_{ij})$, $\forall i \in \{1,...,N\}$, $\forall j \in \{hp\}_i$
\State ${T} \gets \textsf{Add}(T, T_i)$, ${A} \gets \textsf{Add}(A, A_{i})$, $\forall i \in \{1,...,N\}$
\State ${HP_j} \gets \textsf{Divide}(HP_{j}, T)$, $\forall j \in \{hp\}$
\State ${A} \gets \textsf{Divide}(A, T)$
\State Send ${HP_j}$, $\forall j \in \{hp\}$, $A$ to the clients
\State ${HP_j} \gets \textsf{DDecrypt}(HP_j, \{sk_i\})$
\State sort ${HP_j}$ w.r.t. $A$

\end{algorithmic}
\end{algorithm}

\subsubsection{Extensions}\label{sec:extensions} We demonstrated how \sys can be used to implement two privacy-preserving HP tuning strategies, namely, the PF-Mean and PF-DBSCAN. However, it is worth noting that \sys can support other single-shot HP tuning strategies for cross-silo federated learning, provided that their operations can be homomorphically executed at the server side or they can be enabled in a federated manner with assistance from the clients. For example, FLoRA~\cite{flora} proposes several approaches for FL hyperparameter tuning, that are based on performing regression on the (HP, validation accuracy) pairs discovered by each client. Froelicher et al.~\cite{spindle} have shown how such regression operations can be enabled in the federated setting using MHE and \sys, with its approximated functionalities, e.g., division and comparison, can further support the multiple approaches (e.g., MPLM or APLM -- see Section~\ref{sec:comparison-flora}) proposed in FLoRA. In summary, we are confident that \sys can enhance the privacy of various HP tuning methods that rely on finding optimal HPs at the client side and combining them on the server side.

\subsection{Experimental Evaluation}\label{sec:experiments-privtuna}

We experimentally evaluate \sys with a focus on the runtime performance, communication overhead, and (HP tuning) precision, of the private federated mean (PF-Mean) and DBSCAN (PF-DBSCAN). We implement \sys on top of the Lattigo~\cite{lattigo_v5} lattice-based (M)HE library that supports the cryptographic operations.  All experiments are executed on an Apple M2 Pro processor running at 3.49GHz with 16GB of RAM. We conduct our experiments with two HPs, i.e., server learning rate and momentum, and we simulate cross-silo federated learning settings with $N=10$ and $N=20$ clients and all the skew settings and datasets of Section~\ref{sec:design}. We report timings for ring dimensions of both $\mathcal{N}=2^{14}$ and $\mathcal{N}=2^{15}$, which yield an initial level (i.e., the depth of the circuit that can be evaluated before bootstrapping) of $L=10$ and $L=18$, respectively. The ciphertext moduli are $Q\sim2^{438}$ and $Q\sim2^{880}$. These parameters achieve 128-bit security according to the HE standard~\cite{HEStandardPaper}. Finally, for the density-based clustering experiments, we set the grid cell granularity to $L=0.15$ and $\textsf{MinPts}=4$, except for the quantity skew setting, where $\textsf{MinPts}=2$.

\subsubsection{Runtime Performance}

We report total runtime results for two scenarios: one with 10 clients and $\mathcal{N}=2^{14}$ (Table~\ref{table:10clientsN14}), and another with 20 clients and $\mathcal{N}=2^{15}$ (Table~\ref{table:20clientsN15}), without accounting for communication delays. The larger ring size is employed to address the cryptographic noise accumulation that occurs with more clients. Note that the runtime of \sys is independent of the number of HPs as it packs them in the same ciphertext and leverages on SIMD operations. \sys executes PF-Mean among 10 clients in $1.6$s ($\mathcal{N}=2^{14}$) and among 20 clients in $1.8$s ($\mathcal{N}=2^{15}$). The total \sys runtime for PF-DBSCAN among 10 clients is $\sim$5.8s ($\mathcal{N}=2^{14}$). For $N=20$ clients and larger cryptographic parameters ($\mathcal{N}=2^{15}$), \sys runs PF-DBSCAN in $\sim$18.3s. Note that the timings do not include LHO as this process is performed on cleartext data and optimizations on the local HP search are out of the scope of this work. Tables~\ref{table:10clientsN14} and~\ref{table:20clientsN15} provide a more detailed microbenchmark of the cryptographic operations. We observe that $\textsf{DKeyGen}(\{sk_i\})$ and $\textsf{Compare}(\bm{c}_{pk},\bm{t}_{pk})$ are the most time-consuming operations. However, note that $\textsf{DKeyGen}(\{sk_i\})$ is executed only once and that $\textsf{Compare}(\bm{c}_{pk},\bm{t}_{pk})$ relies on an approximation with a circuit of large depth, necessitating the use of the $\textsf{DBootstrap}$ operation to refresh the ciphertexts.


\begin{table}[t]
\caption{Microbenchmarks with $N$=10 clients, $\mathcal{N}=2^{14}$}
\centering
\begin{tabular}{||c|c||}
\hline
Operation                                 & Mean ± Std Dev [ms] \\ \hline
$\textsf{SecKeyGen}(1^\lambda)$ (per client)          & 14.42 ± 1.12    \\
$\textsf{DKeyGen}(\{sk_i\})$ ($pk$ generation)      & 49.27 ± 3.79    \\
$\textsf{DKeyGen}(\{sk_i\})$ ($[ek]$ generation) & 872.18 ± 14.38  \\
$\textsf{Encrypt}(pk,  {p})$                                & 180.94 ± 9.41   \\
$\textsf{Add}(\bm{c}_{pk}, \bm{c'}_{pk})$                      & 4.92 ± 1.57     \\
$\textsf{Mul}_{\textsf{ct}}$($\bm{c}_{pk}, \bm{c'}_{pk})$              & 3.73 ± 1.33     \\
$\textsf{Divide}(\bm{c}_{pk},\bm{c'}_{pk})$                    & 135.69 ± 6.54   \\
$\textsf{DBootstrap}$ (protocol initialization)         & 91.42 ± 4.62    \\
$\textsf{Compare}(\bm{c}_{pk},\bm{t}_{pk})$ (with bootstrapping)   & 3474.72 ± 371.59 \\
$\textsf{DDecrypt}(\bm{c}, \{sk_i\})$                                 & 0.31 ± 0.06   \\
\hline
\end{tabular}
\label{table:10clientsN14}
\end{table}

\begin{table}[t]
\caption{Microbenchmarks with $N$=20 clients, $\mathcal{N}=2^{15}$}
\centering
\begin{tabular}{||c|c||}
\hline
Operation                                 & Mean ± Std Dev [ms] \\ \hline
$\textsf{SecKeyGen}(1^\lambda)$ (per client)          & 51.24 ± 2.41    \\
$\textsf{DKeyGen}(\{sk_i\})$ ($pk$ generation)            & 370.14 ± 3.09    \\
$\textsf{DKeyGen}(\{sk_i\})$ ($[ek]$ generation) & 842.7 ± 5.06  \\
$\textsf{Encrypt}(pk,  {p})$                                 & 189.637 ± 0.64   \\
$\textsf{Add}(\bm{c}_{pk}, \bm{c'}_{pk})$                    & 71.024 ± 12.55   \\
$\textsf{Mul}_{\textsf{ct}}$($\bm{c}_{pk}, \bm{c'}_{pk})$                & 3.49 ± 1.27     \\
$\textsf{Divide}(\bm{c}_{pk},\bm{c'}_{pk})$                      & 154.810 ± 4.22   \\
$\textsf{DBootstrap}$ (protocol initialization)         & 185.85 ± 4.39    \\
$\textsf{Compare}(\bm{c}_{pk},\bm{t}_{pk})$ (with bootstrapping)   & 5021.61 ± 138.93 \\
$\textsf{DDecrypt}(\bm{c}, \{sk_i\})$                                   & 0.29 ± 0.06   \\
\hline
\end{tabular}
\label{table:20clientsN15}
\end{table}

\subsubsection{Communication Overhead.} The communication overhead of \sys depends on the $\textsf{Combine}(\cdot)$ strategy, and the number of clients. A single ciphertext with $\mathcal{N}=2^{14}$ ($\mathcal{N}=2^{15}$) has a size of $2.62$MB ($9.43$MB). We observe that the communication scales linearly with the number of clients. In PF-Mean, each client sends one ciphertext of size $2.62$MB encrypting the sum of the learning rate and momentum discovered by LHO, and another one that encrypts the number of attributed HP sets, yielding a total of $5.24$MB per client and $52.4$MB for 10 clients. In PF-DBSCAN, each client sends 3 ciphertexts (one for the Aggregation-Round 1 and two for Aggregation-Round 2), yielding a total of $7.86$MB and $28.29$MB per client without bootstrapping, for ($\mathcal{N}=2^{14}, N=10$) and ($\mathcal{N}=2^{15}, N=20$), respectively. Finally, the $\textsf{DBootstrap()}$ operation, required for the comparison operations, incurs a communication cost of $\sim13.1MB$ with $\mathcal{N}=2^{14}$, $N=10$ clients. While PF-Mean does not utilize bootstrapping, PF-DBSCAN uses it 6 times. Note that the number of bootstrapping operations depends on the circuit depth and the degree of the approximations; less/more precise implementations require less/more bootstrapping operations.

\subsubsection{Precision.} We evaluate the precision of \sys for HP tuning in the context of both PF-Mean and PF-DBSCAN. We quantify the distortion between the HPs discovered by the plaintext and encrypted federated versions using the mean squared error (MSE). PF-Mean incurs an MSE of $\sim$$2.4\times10^{-4}$ for ($\mathcal{N}=2^{14}, N=10$) and $\sim$$1.8\times10^{-4}$ for ($\mathcal{N}=2^{15}, N=20$) which is mostly due to the division algorithm. PF-DBSCAN, which involves an approximation of both the division and comparison operations, yields an MSE of $\sim$$3.2 \times 10^{-4}$ and $\sim$$1.02 \times 10^{-3}$, for ($\mathcal{N}=2^{14}, N=10$) and ($\mathcal{N}=2^{15}, N=20$), respectively, underscoring the minimal distortion in tuning performance incurred by our framework.

\section{Related Work}\label{sec:related}
We review the related work on HP tuning for both centralized and decentralized machine learning settings and discuss the privacy aspects of HPs.





\descr{Centralized Hyperparameter Tuning.} HP tuning is well-studied in the context of centralized data and several techniques such as grid search~\cite{LeCun2012,lecun2000,Bellman1961}, random search~\cite{bergstra12a}, Bayesian~\cite{Mockus} or evolutionary optimization~\cite{Bardanet}, have been proposed~\cite{Bergstra2011}. While these solutions are considered the norm for HP tuning in centralized machine learning, it is challenging to adapt them to FL because of their iterative nature. In particular, these techniques require multiple rounds of model training, hence yielding a significant communication overhead in FL given its collaborative training process. This overhead is exacerbated when privacy-enhancing technologies such as MPC or HE are deployed to strengthen the privacy of FL.




\descr{Decentralized HP Tuning.} Several works focus on efficient neural architecture search (NAS) solutions tailored to federated learning (FL). Khodak et al. propose several weight-sharing techniques for FL hyperparameter tuning~\cite{khodak2019weight,khodak2021federated}. However, recent literature has shown that weight sharing in FL raises significant privacy concerns~\cite{Hitaj2017,Wang2019,Zhu2019,Melis2019,Nasr2019,Zhao2020,Jonas2020,Wainakh2022}. Other works propose performing NAS offline, i.e., before the federated training starts, or online, i.e., the HP optimization and the training are done simultaneously~\cite{Zhu2021Nas}. Examples of offline NAS solutions are those from Zhu and Ji which propose a novel evolutionary algorithm for NAS in FL~\cite{zhu2019multiobjective} and from Xu et al. which reduce communication by choosing a subset of clients that perform various architecture searches and by eliminating models with low validation errors~\cite{xu2020federated}. However, offline NAS approaches for FL are very costly and require a high number of communication rounds as discussed in~\cite{Zhu2021Nas}. In online federated NAS solutions~\cite{he2021noniid,zhu2020realtime}, all candidate models are trained simultaneously on the clients' side introducing significant computation overhead. We refer interested readers to~\cite{Zhu2021Nas} for a detailed overview of approaches for NAS in the FL setting and we note that contrary to these approaches, our focus is on efficient methods for tuning the FL server hyperparameters and not the model architecture.


To tune FL hyperparameters, such as the number of training passes or the number of clients, Zhang et al. propose FedTune that automatically adjusts them during model training, while respecting any application preferences~\cite{zhang2022fedtune}. Agrawal et al. present a hybrid algorithm that clusters the FL clients based on their hyperparameters which are then updated via a genetic algorithm. Chen et al. propose an online (\textit{tuning while training}) method for both client and server HPs and Dai et al. adapt Bayesian optimization to FL by employing Federated Thompson sampling (FTS)~\cite{Dai2020}. Mostofa describes a representation matching scheme that reduces the divergence of local models from the global one and proposes an online, adaptive HP tuning scheme~\cite{mostafa2020robust}. Kuo et al. study the effect of noisy evaluation in FL HP tuning, and identify data heterogeneity and privacy as key challenges~\cite{kuo2023noisy}. Moreover, they suggest to perform HP tuning on public proxy data for efficiency. Although the aforementioned solutions pioneer HP tuning in FL settings, they are not suitable for privacy-preserving FL solutions; to tune HPs they require access to the federated model which is by design protected with PETs such as MPC or HE, in settings with strong privacy requirements. Moreover, they introduce significant communication and computation overhead which is exacerbated in the presence of PETs. 

To address the communication efficiency issue of prior FL HP tuning approaches, Zhou et al. propose FLoRA, an single-shot HP tuning approach~\cite{flora}. FLoRa discovers the optimal server HPs based on the local HPs of the clients by projecting them to a unified loss surface using various techniques (see Section~\ref{sec:comparison-flora}). While similar to our work, FLoRA differs in two ways: (i) it employs as ground truth the default HP set of scikit-learn algorithms while we construct a more realistic baseline through federated grid search, and (ii) FLoRA does not prevent information leakage from the shared HPs, while we propose a novel framework for privacy-preserving hyperparameter tuning using multi-party homomorphic encryption.

\descr{Hyperparameters and Privacy.} Papernot and Steinke demonstrate that the optimal hyperparameters discovered on a dataset can leak information about its samples via membership inference attacks~\cite{papernot2022hyperparameter}. While the authors propose the usage of Renyi Differential Privacy to eliminate such leakage, applying differentially private techniques during HP tuning can severely degrade utility and introduce further challenges for tuning the DP parameters themselves. Consequently, Koskela and Kulkarni propose using only a random subset of the confidential data for HP tuning in order to achieve lower DP bounds and yield better privacy-utility trade-offs~\cite{koskela2023practical}. However, a recent study applying auditing techniques reveals that the privacy bounds of previous DP-based HP tuning are not tight~\cite{xiang2024does}. Mohapatra et al. investigate the optimization of HPs in the context of DP and show how the HP tuning process can be integrated within the broader privacy budget allocation through adaptive optimizations~\cite{Mohapatra2022}. Ding et al. develop a DP-HP tuning framework with constant privacy overhead, allowing the expansion of the HP search space without affecting the privacy loss, i.e., the privacy loss is independent of the number of HP candidates and the privacy parameter~\cite{Ding2022}. All these studies, however, do not take the FL setting into consideration.

\descr{Privacy-Preserving Hyperparameter Tuning for FL.} Finally, we review works that, similar to ours, aim at protecting data privacy during the hyperparameter tuning process in FL. Zawad et al.~\cite{zawad2022demystifying} propose a proxy-based HP scheme that allows the tuning tasks to be executed on the server side using synthetic data. Their method, however, is tailored to client-side parameters and sharing synthetic data raises additional privacy concerns~\cite{stadler2022synthetic,giomi2023unified}. Seng et al. describe FEATHERS, a novel method for simultaneously performing NAS and HP tuning using DP~\cite{seng2023feathers}. However, FEATHERS operates in an online manner introducing additional communication overhead compared to single-shot HP tuning approaches.




\section{Conclusion}\label{sec:conclusion}


We investigated the open problem of privacy-preserving hyperparameter (HP) tuning in the context of cross-silo federated learning (FL). We first benchmarked various HP tuning strategies through a comprehensive measurement study, on various datasets, model architectures, and FL settings, aiming to understand the relationship between the clients and the server HPs, in both iid and non-iid data settings. Our results allowed us to devise efficient methods based on aggregation and density-based clustering, suitable for deriving the optimal server HPs from the locally computed HPs at individual clients. Aiming to reduce the privacy leakage from the hyperparameter tuning process, we introduced \sys, a novel framework based on multi-party homomorphic encryption that enables the private tuning of HPs in cross-silo FL settings with strong confidentiality requirements. We implemented two strategies from our measurement study, PF-Mean, and PF-DBSCAN, within \sys and we experimentally demonstrated their efficiency and accuracy for federated HP tuning. As future work, we plan to evaluate our methods in settings and tasks with different types of datasets and to explore the extension of our framework with efficient solutions for neural architecture search that can further reduce the costs of privacy-preserving FL pipelines.


\section*{Acknowledgment}
This work was partly supported by the Horizon Europe project HARPOCRATES (Grant agreement 101069535).

\bibliographystyle{abbrv}
\bibliography{bibfile}

\begin{thebibliography}{10}

\bibitem{abadi2016deep}
M.~Abadi, A.~Chu, I.~Goodfellow, H.~B. McMahan, I.~Mironov, K.~Talwar, and L.~Zhang.
\newblock Deep learning with differential privacy.
\newblock In {\em ACM CCS}, 2016.

\bibitem{Agrawal2021}
S.~Agrawal, S.~Sarkar, M.~Alazab, P.~K.~R. Maddikunta, T.~R. Gadekallu, and Q.-V. Pham.
\newblock Genetic cfl: Hyperparameter optimization in clustered federated learning.
\newblock {\em Computational Intelligence and Neuroscience}, 2021:7156420, Nov 2021.

\bibitem{HEStandardPaper}
M.~Albrecht et~al.
\newblock Homomorphic {E}ncryption {S}ecurity {S}tandard.
\newblock Technical report, HomomorphicEncryption.org, 2018.

\bibitem{Bardanet}
R.~Bardenet and B.~Kégl.
\newblock Surrogating the surrogate: accelerating gaussian-process-based global optimization with a mixture cross-entropy algorithm.
\newblock 08 2010.

\bibitem{Bellman1961}
R.~E. Bellman.
\newblock {\em Adaptive Control Processes: A Guided Tour}.
\newblock Princeton University Press, Princeton, 1961.

\bibitem{Bergstra2011}
J.~Bergstra, R.~Bardenet, B.~Kégl, and Y.~Bengio.
\newblock Algorithms for hyper-parameter optimization.
\newblock 12 2011.

\bibitem{bergstra12a}
J.~Bergstra and Y.~Bengio.
\newblock Random search for hyper-parameter optimization.
\newblock {\em Journal of Machine Learning Research}, 13(10):281--305, 2012.

\bibitem{Bonawitz2016}
K.~Bonawitz, V.~Ivanov, B.~Kreuter, A.~Marcedone, H.~B. McMahan, S.~Patel, D.~Ramage, A.~Segal, and K.~Seth.
\newblock Practical secure aggregation for federated learning on user-held data.
\newblock In {\em NIPS PPML Workshop}, 2016.

\bibitem{briggs2020federated}
C.~Briggs, Z.~Fan, and P.~Andras.
\newblock Federated learning with hierarchical clustering of local updates to improve training on non-iid data.
\newblock In {\em 2020 International Joint Conference on Neural Networks (IJCNN)}, pages 1--9. IEEE, 2020.

\bibitem{flash}
M.~Byali, H.~Chaudhari, A.~Patra, and A.~Suresh.
\newblock {FLASH: F}ast and robust framework for privacy-preserving machine learning.
\newblock {\em PETS}, 2020.

\bibitem{trident}
H.~Chaudhari, R.~Rachuri, and A.~Suresh.
\newblock {Trident: E}fficient 4pc framework for privacy preserving machine learning.
\newblock In {\em Network and Distributed System Security Symposium (NDSS)}, 2020.

\bibitem{Chen}
T.~{Chen} and S.~{Zhong}.
\newblock Privacy-preserving backpropagation neural network learning.
\newblock {\em IEEE Transactions on Neural Networks}, 20(10):1554--1564, Oct 2009.

\bibitem{chen2019distributed}
X.~Chen, T.~Chen, H.~Sun, Z.~S. Wu, and M.~Hong.
\newblock Distributed training with heterogeneous data: Bridging median- and mean-based algorithms.
\newblock {\em CoRR}, abs/1906.01736, 2019.

\bibitem{cheon2017homomorphic}
J.~H. Cheon, A.~Kim, M.~Kim, and Y.~Song.
\newblock Homomorphic encryption for arithmetic of approximate numbers.
\newblock In {\em ASIACRYPT}, 2017.

\bibitem{Cheon2020}
J.~H. Cheon, D.~Kim, and D.~Kim.
\newblock Efficient homomorphic comparison methods with optimal complexity.
\newblock In S.~Moriai and H.~Wang, editors, {\em Advances in Cryptology -- ASIACRYPT 2020}, pages 221--256, Cham, 2020. Springer International Publishing.

\bibitem{cohen2017emnist}
G.~Cohen, S.~Afshar, J.~Tapson, and A.~Van~Schaik.
\newblock Emnist: Extending mnist to handwritten letters.
\newblock In {\em 2017 International Joint Conference on Neural Networks (IJCNN)}, pages 2921--2926. IEEE, 2017.

\bibitem{corrigan2017prio}
H.~Corrigan-Gibbs and D.~Boneh.
\newblock Prio: {P}rivate, {R}obust, and {C}omputation of {A}ggregate {S}tatistics.
\newblock In {\em USENIX NSDI}, 2017.

\bibitem{Dai2020}
Z.~Dai, B.~K.~H. Low, and P.~Jaillet.
\newblock Federated bayesian optimization via thompson sampling.
\newblock In {\em NeurIPS}, 2020.

\bibitem{Deng2021}
J.~Deng, Y.~Wang, J.~Li, C.~Wang, C.~Shang, H.~Liu, S.~Rajasekaran, and C.~Ding.
\newblock {TAG}: Gradient attack on transformer-based language models.
\newblock In {\em EMNLP}, pages 3600--3610, 2021.

\bibitem{Dimitrov2022}
D.~I. Dimitrov, M.~Balunović, N.~Jovanović, and M.~Vechev.
\newblock Lamp: Extracting text from gradients with language model priors.
\newblock {\em CoRR}, abs/2202.08827, 2022.

\bibitem{Ding2022}
Y.~Ding and X.~Wu.
\newblock Revisiting hyperparameter tuning with differential privacy.
\newblock {\em CoRR}, abs/2211.01852, 2022.

\bibitem{lattigo_v5}
Lattigo v5.
\newblock Online: \url{https://github.com/tuneinsight/lattigo}, Nov. 2023.
\newblock (Accessed: 2023-11-11).

\bibitem{spindle}
D.~Froelicher, J.~R. Troncoso-Pastoriza, A.~Pyrgelis, S.~Sav, J.~S. Sousa, J.-P. Bossuat, and J.-P. Hubaux.
\newblock Scalable privacy-preserving distributed learning.
\newblock {\em PETS}, 2021.

\bibitem{Froelicher2021}
D.~Froelicher, J.~R. Troncoso-Pastoriza, J.~L. Raisaro, M.~A. Cuendet, J.~S. Sousa, H.~Cho, B.~Berger, J.~Fellay, and J.-P. Hubaux.
\newblock Truly privacy-preserving federated analytics for precision medicine with multiparty homomorphic encryption.
\newblock {\em Nature Communications}, 12(1):5910, Oct 2021.

\bibitem{Jonas2020}
J.~Geiping, H.~Bauermeister, H.~Dr\"{o}ge, and M.~Moeller.
\newblock Inverting gradients - how easy is it to break privacy in federated learning?
\newblock In {\em NeurIPS}, 2020.

\bibitem{giomi2023unified}
M.~Giomi, F.~Boenisch, C.~Wehmeyer, and B.~Tasn{\'a}di.
\newblock A unified framework for quantifying privacy risk in synthetic data.
\newblock {\em Proceedings on Privacy Enhancing Technologies}, 2:312--328, 2023.

\bibitem{Goldschmidt}
R.~Goldschmidt.
\newblock Applications of division by convergence.
\newblock M.sc. dissertation. {M.I.T.}, 1964.

\bibitem{he2021noniid}
C.~He, M.~Annavaram, and S.~Avestimehr.
\newblock Towards non-i.i.d. and invisible data with fednas: Federated deep learning via neural architecture search.
\newblock {\em CoRR}, abs/2004.08546, 2021.

\bibitem{cryptoDL}
E.~Hesamifard, H.~Takabi, M.~Ghasemi, and R.~Wright.
\newblock Privacy-preserving machine learning as a service.
\newblock {\em PETS}, 2018.

\bibitem{DTI}
B.~Hie, H.~Cho, and B.~Berger.
\newblock Realizing private and practical pharmacological collaboration.
\newblock {\em Science}, 362(6412):347--350, 2018.

\bibitem{Hitaj2017}
B.~Hitaj, G.~Ateniese, and F.~Perez-Cruz.
\newblock Deep models under the {GAN}: Information leakage from collaborative deep learning.
\newblock In {\em ACM CCS}, 2017.

\bibitem{jayaraman2018distributed}
B.~Jayaraman, L.~Wang, D.~Evans, and Q.~Gu.
\newblock Distributed learning without distress: Privacy-preserving empirical risk minimization.
\newblock In {\em NIPS}, 2018.

\bibitem{Jiang2018}
P.~Jiang and G.~Agrawal.
\newblock A linear speedup analysis of distributed deep learning with sparse and quantized communication.
\newblock In {\em Proceedings of the 32nd International Conference on Neural Information Processing Systems}, NIPS'18, page 2530–2541, Red Hook, NY, USA, 2018. Curran Associates Inc.

\bibitem{Jin2021}
X.~Jin, P.-Y. Chen, C.-Y. Hsu, C.-M. Yu, and T.~Chen.
\newblock Cafe: Catastrophic data leakage in vertical federated learning.
\newblock In {\em NeurIPS}, volume~34, pages 994--1006, 2021.

\bibitem{kairouz2019}
P.~Kairouz et~al.
\newblock Advances and open problems in federated learning.
\newblock {\em CoRR}, arXiv:1912.04977, 2019.

\bibitem{keller2022secure}
M.~Keller and K.~Sun.
\newblock Secure quantized training for deep learning.
\newblock In {\em International Conference on Machine Learning}, pages 10912--10938. PMLR, 2022.

\bibitem{khodak2019weight}
M.~Khodak, L.~Li, N.~Roberts, M.-F. Balcan, and A.~Talwalkar.
\newblock Weight-sharing beyond neural architecture search: Efficient feature map selection and federated hyperparameter tuning.
\newblock In {\em Proc. 2nd SysML Conf.}, 2019.

\bibitem{khodak2021federated}
M.~Khodak, R.~Tu, T.~Li, L.~Li, N.~Balcan, V.~Smith, and A.~Talwalkar.
\newblock Federated hyperparameter tuning: Challenges, baselines, and connections to weight-sharing.
\newblock In A.~Beygelzimer, Y.~Dauphin, P.~Liang, and J.~W. Vaughan, editors, {\em Advances in Neural Information Processing Systems}, 2021.

\bibitem{knott2021crypten}
B.~Knott, S.~Venkataraman, A.~Hannun, S.~Sengupta, M.~Ibrahim, and L.~van~der Maaten.
\newblock Crypten: Secure multi-party computation meets machine learning.
\newblock {\em Advances in Neural Information Processing Systems}, 34:4961--4973, 2021.

\bibitem{Konency2016fed}
J.~Kone{\v{c}}n{\`y}, H.~B. McMahan, D.~Ramage, and P.~Richt{\'a}rik.
\newblock Federated optimization: Distributed machine learning for on-device intelligence.
\newblock {\em CoRR}, abs:1610.02527, 2016.

\bibitem{Konecny2016}
J.~Konecn{\'{y}}, H.~B. McMahan, F.~X. Yu, P.~Richt{\'{a}}rik, A.~T. Suresh, and D.~Bacon.
\newblock Federated learning: Strategies for improving communication efficiency.
\newblock {\em CoRR}, abs/1610.05492, 2016.

\bibitem{koskela2023practical}
A.~Koskela and T.~Kulkarni.
\newblock Practical differentially private hyperparameter tuning with subsampling.
\newblock {\em CoRR}, arXiv:2301.11989, 2023.

\bibitem{cifarPaper}
A.~Krizhevsky.
\newblock Learning multiple layers of features from tiny images.
\newblock {\em Technical Report, University of Toronto}, 2012.

\bibitem{kuo2023noisy}
K.~Kuo, P.~Thaker, M.~Khodak, J.~Nguyen, D.~Jiang, A.~Talwalkar, and V.~Smith.
\newblock On noisy evaluation in federated hyperparameter tuning.
\newblock In {\em MLSys}, 2023.

\bibitem{lecun2000}
Y.~Lecun, L.~Bottou, G.~Orr, and K.-R. Müller.
\newblock Efficient backprop.
\newblock 08 2000.

\bibitem{MNIST}
Y.~LeCun and C.~Cortes.
\newblock {MNIST} handwritten digit database.
\newblock 2010.

\bibitem{LeCun2012}
Y.~A. LeCun, L.~Bottou, G.~B. Orr, and K.-R. M{\"u}ller.
\newblock {\em Efficient BackProp}, pages 9--48.
\newblock Springer Berlin Heidelberg, Berlin, Heidelberg, 2012.

\bibitem{li2021federated}
Q.~Li, Y.~Diao, Q.~Chen, and B.~He.
\newblock Federated learning on non-iid data silos: An experimental study.
\newblock In {\em 2022 IEEE 38th International Conference on Data Engineering (ICDE)}, pages 965--978, 2022.

\bibitem{Nvidia_Fed}
W.~Li, F.~Milletar{\`i}, D.~Xu, N.~Rieke, J.~Hancox, W.~Zhu, M.~Baust, Y.~Cheng, S.~Ourselin, M.~J. Cardoso, and A.~Feng.
\newblock Privacy-preserving federated brain tumour segmentation.
\newblock In {\em Springer MLMI}, 2019.

\bibitem{marino2022}
G.~Marino.
\newblock Federated dbscan.
\newblock Bachelor's thesis, Università di Pisa, Dipartimento di Ingegneria dell'Informazione, 2022.

\bibitem{federatedLearning1}
H.~B. McMahan, E.~Moore, D.~Ramage, and B.~A. y~Arcas.
\newblock Federated learning of deep networks using model averaging.
\newblock {\em CoRR}, abs/1602.05629, 2016.

\bibitem{mcmahan2018LSTM}
H.~B. McMahan, D.~Ramage, K.~Talwar, and L.~Zhang.
\newblock Learning differentially private recurrent language models.
\newblock {\em CoRR}, abs/1710.06963, 2018.

\bibitem{McMahan2018}
H.~B. McMahan, D.~Ramage, K.~Talwar, and L.~Zhang.
\newblock Learning differentially private recurrent language models.
\newblock In {\em ICLR}, 2018.

\bibitem{Melis2019}
L.~{Melis}, C.~{Song}, E.~{De Cristofaro}, and V.~{Shmatikov}.
\newblock Exploiting unintended feature leakage in collaborative learning.
\newblock In {\em IEEE S\&P}, 2019.

\bibitem{Mockus}
J.~Mockus, V.~Tiesis, and A.~Zilinskas.
\newblock {\em The application of Bayesian methods for seeking the extremum}, volume~2, pages 117--129.
\newblock 09 2014.

\bibitem{Mohapatra2022}
S.~Mohapatra, S.~Sasy, X.~He, G.~Kamath, and O.~Thakkar.
\newblock The role of adaptive optimizers for honest private hyperparameter selection.
\newblock {\em Proceedings of the AAAI Conference on Artificial Intelligence}, 36(7):7806--7813, Jun. 2022.

\bibitem{mostafa2020robust}
H.~Mostafa.
\newblock Robust federated learning through representation matching and adaptive hyper-parameters, 2020.

\bibitem{mouchet2019distributedbfv}
C.~Mouchet, J.~R. Troncoso-pastoriza, J.-P. Bossuat, and J.~P. Hubaux.
\newblock Multiparty homomorphic encryption: From theory to practice.
\newblock In {\em Technical Report \url{https://eprint.iacr.org/2020/304}}, 2019.

\bibitem{Nasr2019}
M.~{Nasr}, R.~{Shokri}, and A.~{Houmansadr}.
\newblock Comprehensive privacy analysis of deep learning: Passive and active white-box inference attacks against centralized and federated learning.
\newblock In {\em IEEE S\&P}, 2019.

\bibitem{svhn}
Y.~Netzer, T.~Wang, A.~Coates, A.~Bissacco, B.~Wu, and A.~Ng.
\newblock Reading digits in natural images with unsupervised feature learning.
\newblock {\em NIPS}, 2011.

\bibitem{Ozkok}
F.~O. Ozkok and M.~Celik.
\newblock A new approach to determine eps parameter of dbscan algorithm.
\newblock {\em International Journal of Intelligent Systems and Applications in Engineering}, 5(4):247–251, Dec. 2017.

\bibitem{papernot2022hyperparameter}
N.~Papernot and T.~Steinke.
\newblock Hyperparameter tuning with renyi differential privacy.
\newblock {\em CoRR}, abs/2110.03620, 2022.

\bibitem{Phong2017}
L.~T. Phong, Y.~Aono, T.~Hayashi, L.~Wang, and S.~Moriai.
\newblock Privacy-preserving deep learning: Revisited and enhanced.
\newblock In {\em Springer ATIS}, 2017.

\bibitem{Phong2018}
L.~T. {Phong}, Y.~{Aono}, T.~{Hayashi}, L.~{Wang}, and S.~{Moriai}.
\newblock Privacy-preserving deep learning via additively homomorphic encryption.
\newblock {\em IEEE TIFS}, 13(5):1333--1345, 2018.

\bibitem{Cock2021}
D.~Reich, A.~Todoki, R.~Dowsley, M.~D. Cock, and A.~C.~A. Nascimento.
\newblock Privacy-preserving classification of personal text messages with secure multi-party computation: An application to hate-speech detection.
\newblock {\em CoRR}, abs:1906.02325, 2021.

\bibitem{fedjax2021}
J.~H. Ro, A.~T. Suresh, and K.~Wu.
\newblock {F}ed{JAX}: Federated learning simulation with {JAX}.
\newblock {\em arXiv preprint arXiv:2108.02117}, 2021.

\bibitem{sander}
J.~Sander, M.~Ester, H.-P. Kriegel, and X.~Xu.
\newblock Density-based clustering in spatial databases: The algorithm gdbscan and its applications.
\newblock {\em Data Mining and Knowledge Discovery}, 2:169--194, 1998.

\bibitem{Sav2022}
S.~Sav, J.-P. Bossuat, J.~R. Troncoso-Pastoriza, M.~Claassen, and J.-P. Hubaux.
\newblock Privacy-preserving federated neural network learning for disease-associated cell classification.
\newblock {\em bioRxiv}, 2022.

\bibitem{rhode}
S.~Sav, A.~Diaa, A.~Pyrgelis, J.-P. Bossuat, and J.-P. Hubaux.
\newblock Privacy-preserving federated recurrent neural networks.
\newblock {\em CoRR}, abs/2207.13947, 2022.

\bibitem{poseidon}
S.~Sav, A.~Pyrgelis, J.~R. Troncoso-Pastoriza, D.~Froelicher, J.-P. Bossuat, J.~S. Sousa, and J.-P. Hubaux.
\newblock Poseidon: Privacy-preserving federated neural network learning.
\newblock In {\em Network and Distributed System Security Symposium (NDSS)}, 2021.

\bibitem{seng2023feathers}
J.~Seng, P.~Prasad, M.~Mundt, D.~S. Dhami, and K.~Kersting.
\newblock Feathers: Federated architecture and hyperparameter search.
\newblock {\em CoRR}, arXiv:2206.12342, 2023.

\bibitem{shokri2015privacy}
R.~Shokri and V.~Shmatikov.
\newblock Privacy-preserving deep learning.
\newblock In {\em ACM Conference on Computer and Communications Security (CCS)}, 2015.

\bibitem{stadler2022synthetic}
T.~Stadler, B.~Oprisanu, and C.~Troncoso.
\newblock Synthetic data--anonymisation groundhog day.
\newblock In {\em 31st USENIX Security Symposium (USENIX Security 22)}, pages 1451--1468, 2022.

\bibitem{truex2019hybrid}
S.~Truex, N.~Baracaldo, A.~Anwar, T.~Steinke, H.~Ludwig, R.~Zhang, and Y.~Zhou.
\newblock A hybrid approach to privacy-preserving federated learning.
\newblock In {\em ACM AISec}, 2019.

\bibitem{wagh2019securenn}
S.~Wagh, D.~Gupta, and N.~Chandran.
\newblock Securenn: 3-party secure computation for neural network training.
\newblock {\em PETS}, 2019.

\bibitem{falcon}
S.~Wagh, S.~Tople, F.~Benhamouda, E.~Kushilevitz, P.~Mittal, and T.~Rabin.
\newblock {FALCON: H}onest-majority maliciously secure framework for private deep learning.
\newblock {\em PETS}, 2020.

\bibitem{Wainakh2022}
A.~Wainakh, F.~Ventola, T.~Müßig, J.~Keim, C.~Garcia~Cordero, E.~Zimmer, T.~Grube, K.~Kersting, and M.~Mühlhäuser.
\newblock User-level label leakage from gradients in federated learning.
\newblock {\em PETS}, 2022:227--244, 04 2022.

\bibitem{wang2023dphypo}
H.~Wang, S.~Gao, H.~Zhang, W.~J. Su, and M.~Shen.
\newblock Dp-hypo: An adaptive private hyperparameter optimization framework, 2023.

\bibitem{wang2019adaptive}
S.~Wang, T.~Tuor, T.~Salonidis, K.~K. Leung, C.~Makaya, T.~He, and K.~Chan.
\newblock Adaptive federated learning in resource constrained edge computing systems.
\newblock {\em CoRR}, arXiv:1804.05271, 2019.

\bibitem{Wang2019}
Z.~{Wang}, M.~{Song}, Z.~{Zhang}, Y.~{Song}, Q.~{Wang}, and H.~{Qi}.
\newblock Beyond inferring class representatives: User-level privacy leakage from federated learning.
\newblock In {\em IEEE INFOCOM}, 2019.

\bibitem{Wei2020}
K.~Wei, J.~Li, M.~Ding, C.~Ma, H.~H. Yang, F.~Farokhi, S.~Jin, T.~Q.~S. Quek, and H.~V. Poor.
\newblock Federated learning with differential privacy: Algorithms and performance analysis.
\newblock {\em IEEE Transactions on Information Forensics and Security}, 15:3454--3469, 2020.

\bibitem{wu2019value}
N.~Wu, F.~Farokhi, D.~Smith, and M.~A. Kaafar.
\newblock The value of collaboration in convex machine learning with differential privacy.
\newblock {\em CoRR}, abs/1906.09679, 2019.

\bibitem{xiang2024does}
Z.~Xiang, C.~Wang, and D.~Wang.
\newblock How does selection leak privacy: Revisiting private selection and improved results for hyper-parameter tuning.
\newblock {\em CoRR}, abs:2402.13087, 2024.

\bibitem{xu2020federated}
M.~Xu, Y.~Zhao, K.~Bian, G.~Huang, Q.~Mei, and X.~Liu.
\newblock Federated neural architecture search.
\newblock {\em CoRR}, arXiv:2002.06352, 2020.

\bibitem{yin2021byzantinerobust}
D.~Yin, Y.~Chen, K.~Ramchandran, and P.~Bartlett.
\newblock Byzantine-robust distributed learning: Towards optimal statistical rates.
\newblock {\em CoRR}, abs/1803.01498, 2021.

\bibitem{yu2019linear}
H.~Yu, R.~Jin, and S.~Yang.
\newblock On the linear speedup analysis of communication efficient momentum sgd for distributed non-convex optimization.
\newblock {\em CoRR}, abs/1905.03817, 2019.

\bibitem{zawad2022demystifying}
S.~Zawad, J.~Yi, M.~Zhang, C.~Li, F.~Yan, and Y.~He.
\newblock Demystifying hyperparameter optimization in federated learning, 2022.

\bibitem{Zhang20202}
C.~Zhang, S.~Li, J.~Xia, W.~Wang, F.~Yan, and Y.~Liu.
\newblock Batchcrypt: Efficient homomorphic encryption for cross-silo federated learning.
\newblock In {\em Proceedings of the 2020 USENIX Conference on Usenix Annual Technical Conference}, USENIX ATC'20, USA, 2020. USENIX Association.

\bibitem{zhang2022fedtune}
H.~Zhang, M.~Zhang, X.~Liu, P.~Mohapatra, and M.~DeLucia.
\newblock Fedtune: Automatic tuning of federated learning hyper-parameters from system perspective.
\newblock {\em CoRR}, abs/2110.03061, 2022.

\bibitem{Zhao2020}
B.~Zhao, K.~R. Mopuri, and H.~Bilen.
\newblock i{DLG}: Improved deep leakage from gradients.
\newblock {\em CoRR}, abs/2001.02610, 2020.

\bibitem{zheng2019helen}
W.~Zheng, R.~A. Popa, J.~E. Gonzalez, and I.~Stoica.
\newblock Helen: Maliciously secure coopetitive learning for linear models.
\newblock In {\em IEEE S\&P}, 2019.

\bibitem{flora}
Y.~Zhou, P.~Ram, T.~Salonidis, N.~B. Angel, H.~Samulowitz, and H.~Ludwig.
\newblock Single-shot general hyper-parameter optimization for federated learning.
\newblock In {\em International Conference on Learning Representations}, 2023.

\bibitem{zhu2019multiobjective}
H.~Zhu and Y.~Jin.
\newblock Multi-objective evolutionary federated learning.
\newblock {\em CoRR}, arXiv:1812.07478, 2019.

\bibitem{zhu2020realtime}
H.~Zhu and Y.~Jin.
\newblock Real-time federated evolutionary neural architecture search.
\newblock {\em CoRR}, arXiv:2003.02793, 2020.

\bibitem{Zhu2020}
H.~Zhu, R.~S. Mong~Goh, and W.-K. Ng.
\newblock Privacy-preserving weighted federated learning within the secret sharing framework.
\newblock {\em IEEE Access}, 8:198275--198284, 2020.

\bibitem{Zhu2021Nas}
H.~Zhu, H.~Zhang, and Y.~Jin.
\newblock From federated learning to federated neural architecture search: a survey.
\newblock {\em Complex \& Intelligent Systems}, 7, 01 2021.

\bibitem{Zhu2019}
L.~Zhu, Z.~Liu, and S.~Han.
\newblock Deep leakage from gradients.
\newblock In {\em NeurIPS}, volume~32. 2019.

\end{thebibliography}

\newcounter{subsubsubsection}
\appendices
\subsection{Density-based Clustering}\label{subsec:dbscan}

\begin{figure*}[t]
\centering
    \begin{subfigure}[c]{\columnwidth}
    \centering
    \includegraphics[width=\textwidth]{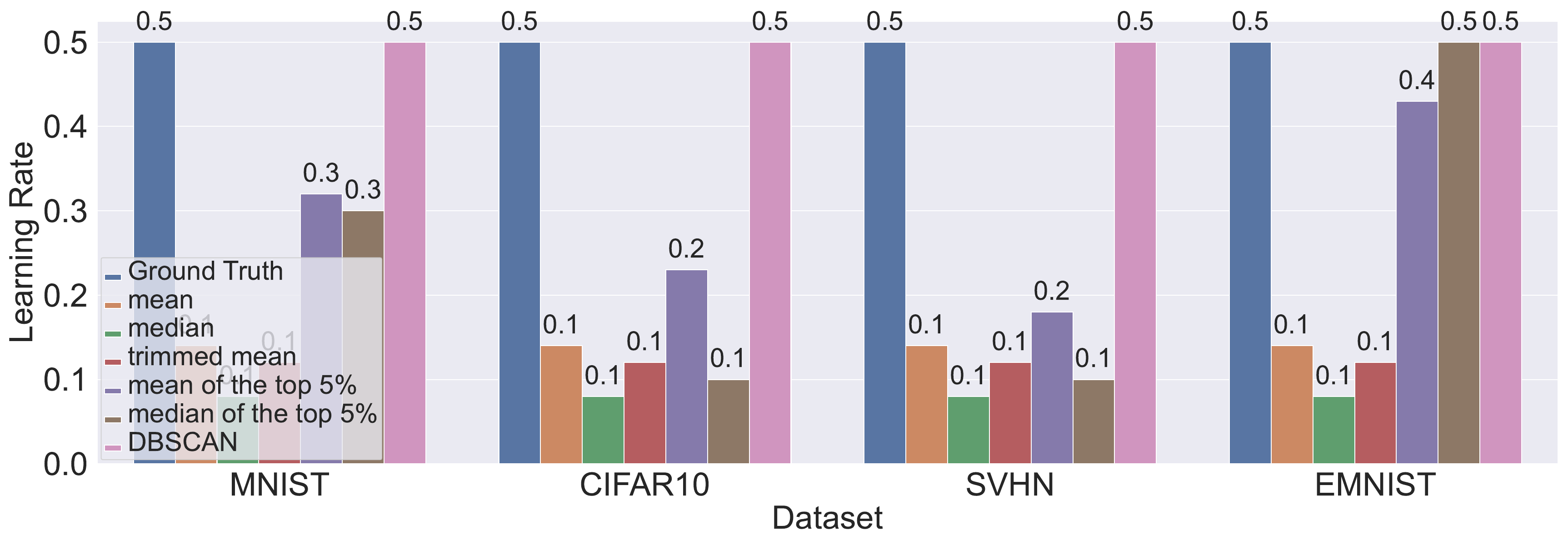} 
    \captionsetup{width=\textwidth}
    \caption{Learning rate with feature skew ($\beta_f=0.1$)}
    \label{fig:measurement_feature_lr-2}
    \end{subfigure}
    \begin{subfigure}[c]{\columnwidth}
    \centering
    \includegraphics[width=\textwidth]{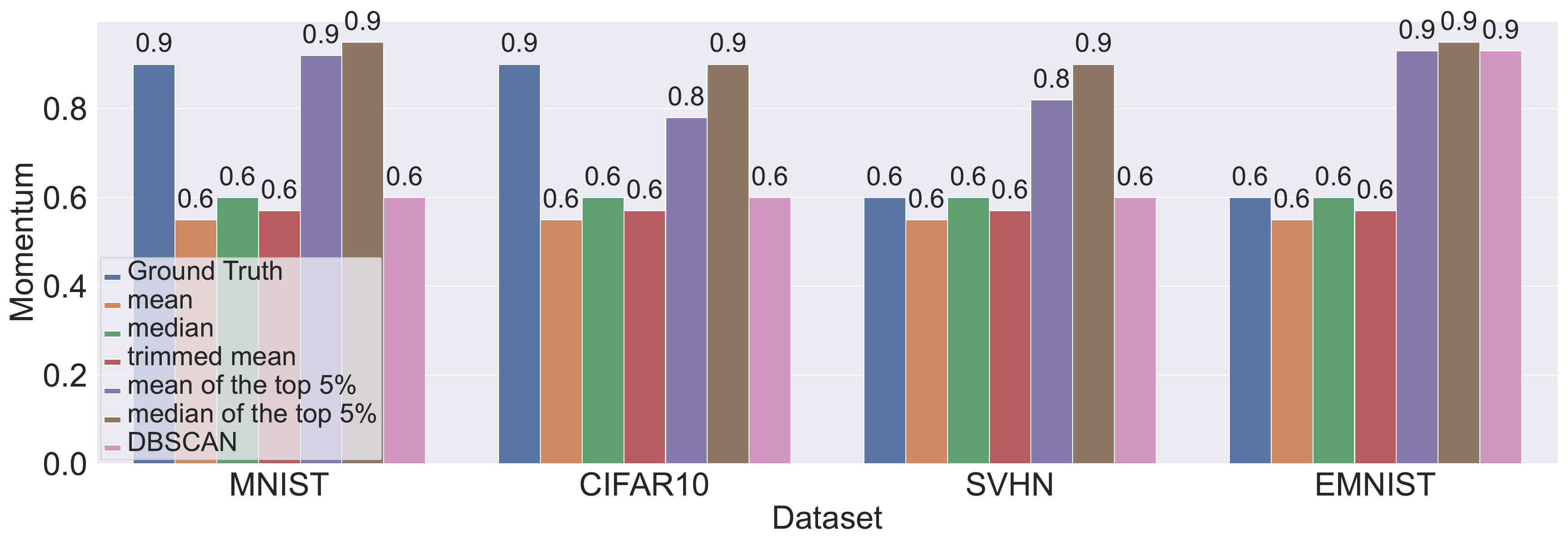}
    \captionsetup{width=\textwidth}
    \caption{Momentum with feature skew ($\beta_f=0.1$)}
    \label{fig:measurement_feature_mom-2}
    \end{subfigure}
    \begin{subfigure}[c]{\columnwidth}
    \centering
    \includegraphics[width=\textwidth]{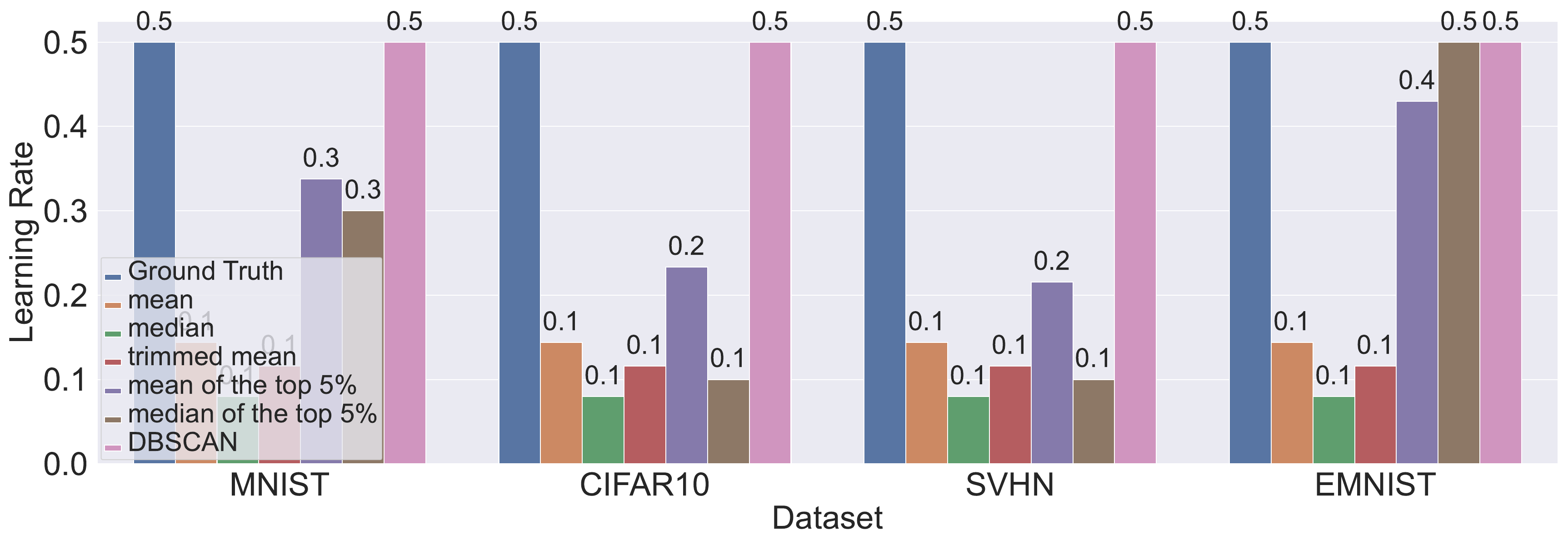} 
    \captionsetup{width=\textwidth}
    \caption{Learning rate with label skew ($\beta_{\ell}=5.0$)}
    \label{fig:measurement_label_lr-2}
    \end{subfigure}
    \begin{subfigure}[c]{\columnwidth}
    \centering
    \includegraphics[width=\textwidth]{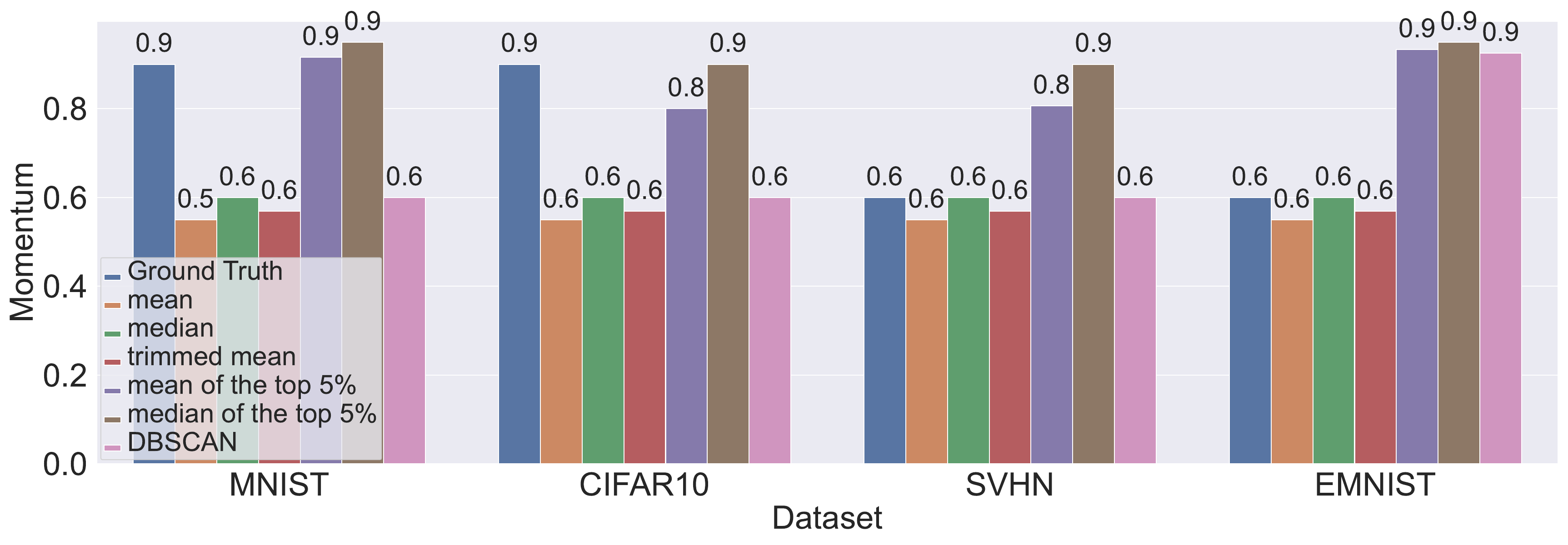} 
    \captionsetup{width=\textwidth}
    \caption{Momentum with label skew ($\beta_{\ell}=5.0$)}
    \label{fig:measurement_label_mom-2}
    \end{subfigure}
    \begin{subfigure}[c]{\columnwidth}
    \centering
    \includegraphics[width=\textwidth]{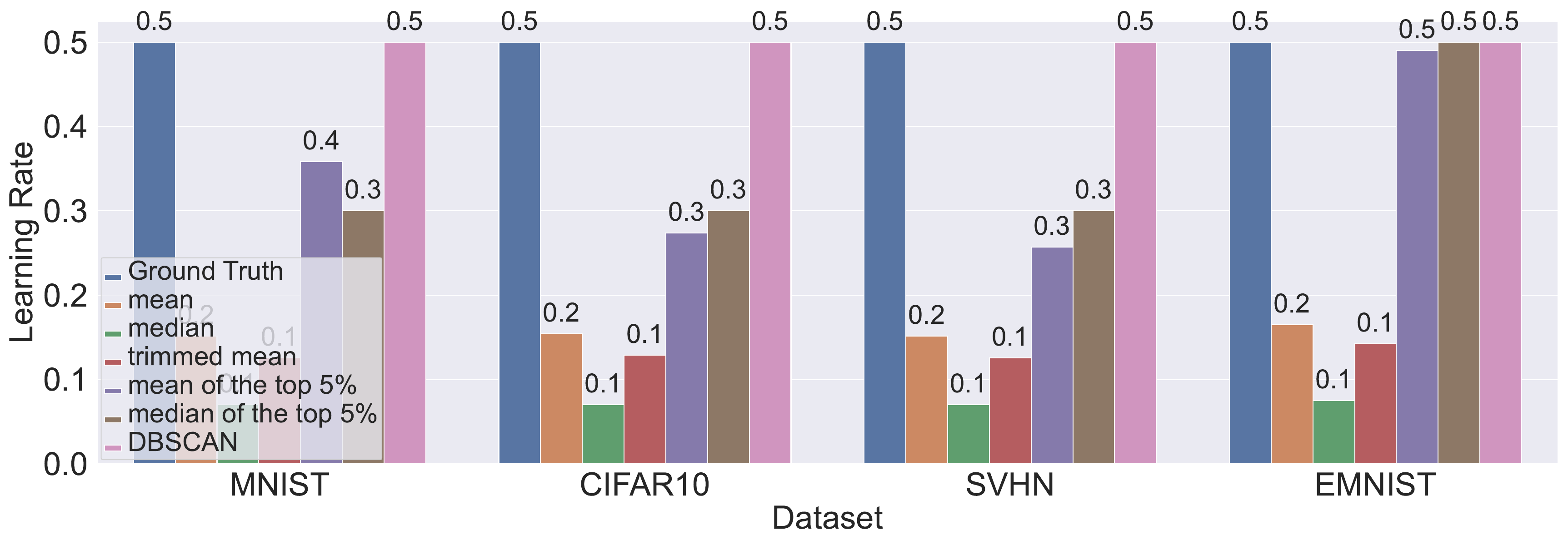} 
    \captionsetup{width=\textwidth}
    \caption{Learning rate with quantity skew ($\beta_q=2.0$)}
    \label{fig:measurement_qty_lr-2}
    \end{subfigure}
    \begin{subfigure}[c]{\columnwidth}
    \centering
    \includegraphics[width=\textwidth]{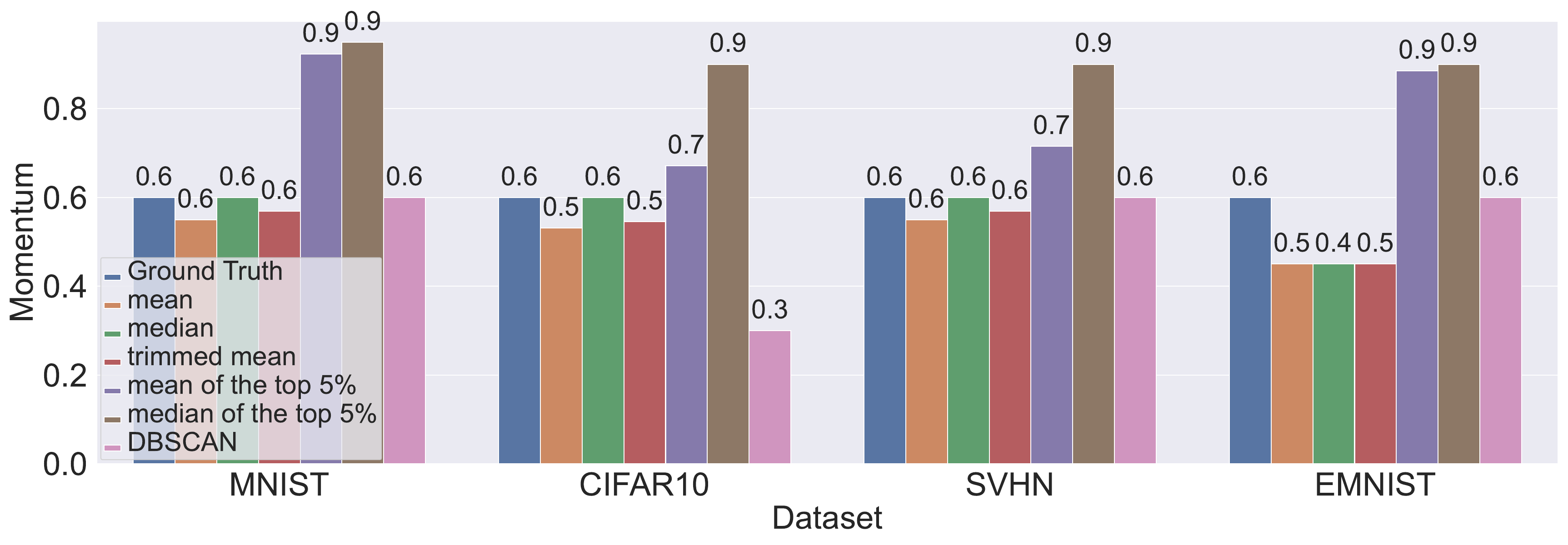}
    \captionsetup{width=\textwidth}
    \caption{Momentum with quantity skew ($\beta_q=2.0$)}
    \label{fig:measurement_qty_mom-2}
    \end{subfigure}
    \caption{Barplots of learning rate and momentum for the non-iid setting with $N=20$ clients, variable datasets and $\textsf{Combine}(\cdot)$ strategies. The top-row results are for feature skew ($\beta_f=0.1$), middle-row results for label skew ($\beta_{\ell}=5.0$), and bottom-row results are for quantity skew ($\beta_q=2.0$). Bar colors represent the hyperparameters derived using the various combination strategies.}
    \label{fig:measurement-2}
\end{figure*}

\begin{figure*}[t]
\centering
    \begin{subfigure}[b]{.33\textwidth}
    \centering
    \includegraphics[width=\textwidth]{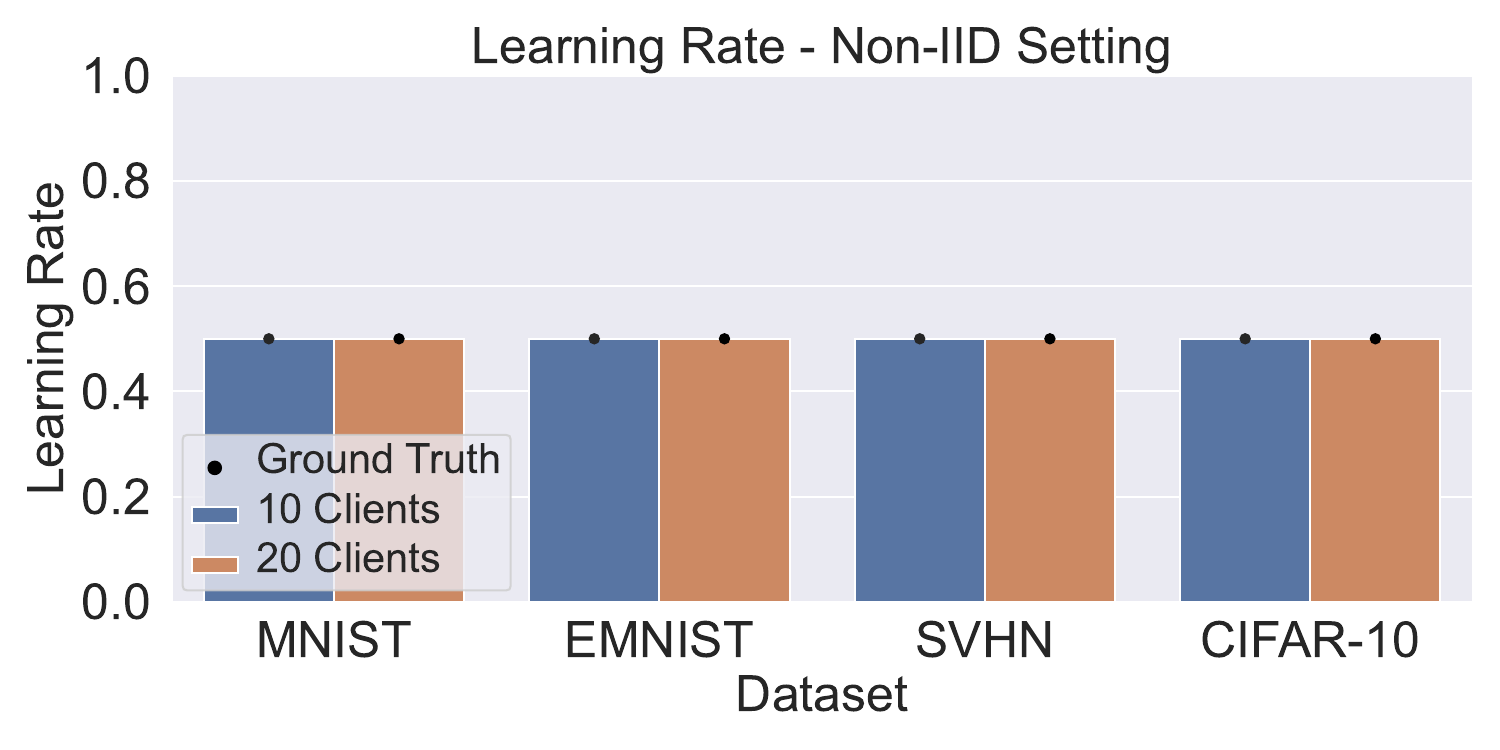}
    \captionsetup{width=\textwidth}
    \caption{Learning Rate}
    \end{subfigure}
    \hfill
    \begin{subfigure}[b]{.33\textwidth}
    \centering
    \includegraphics[width=\textwidth]{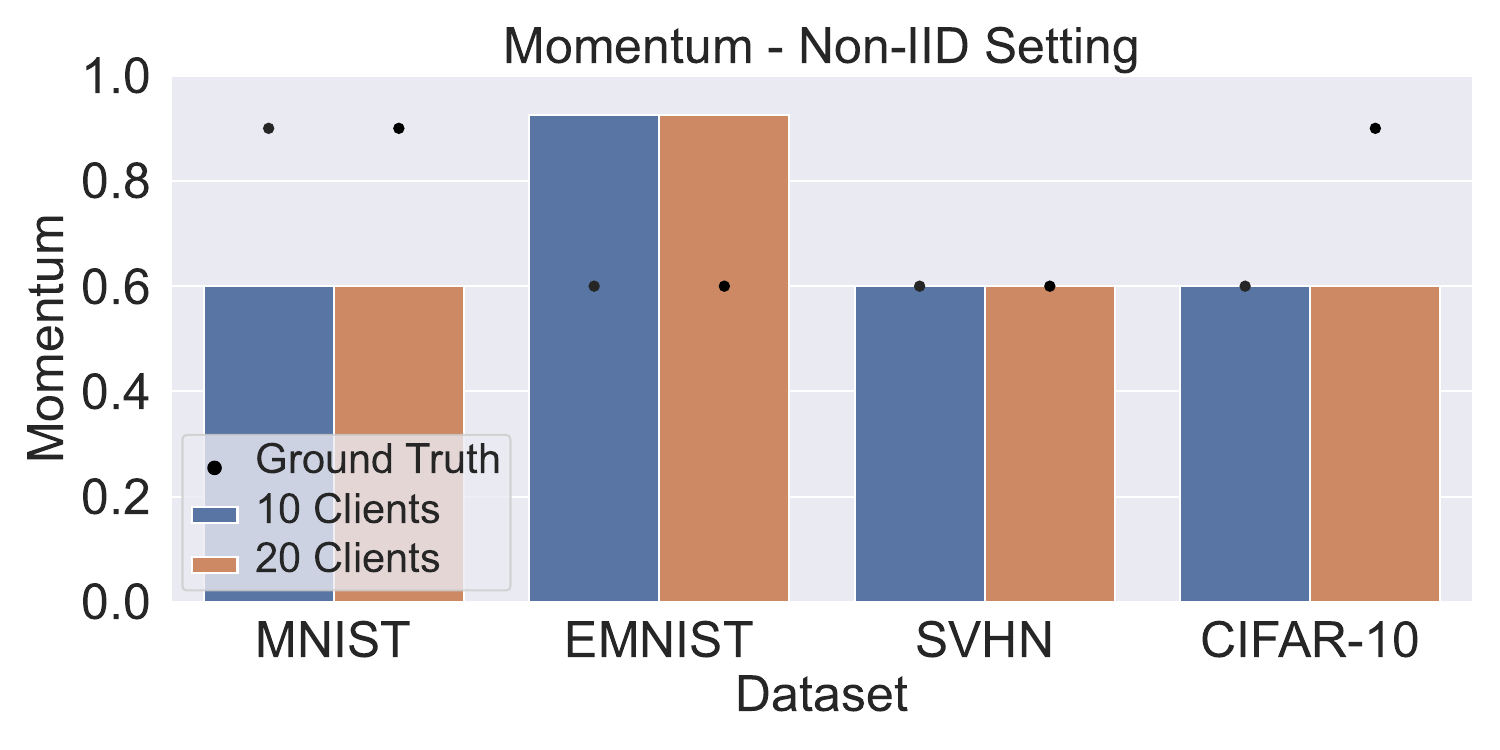}
    \captionsetup{width=\textwidth}
    \caption{Momentum}
    \end{subfigure}
    \hfill
    \begin{subfigure}[b]{.33\textwidth}
    \centering
    \includegraphics[width=\textwidth]{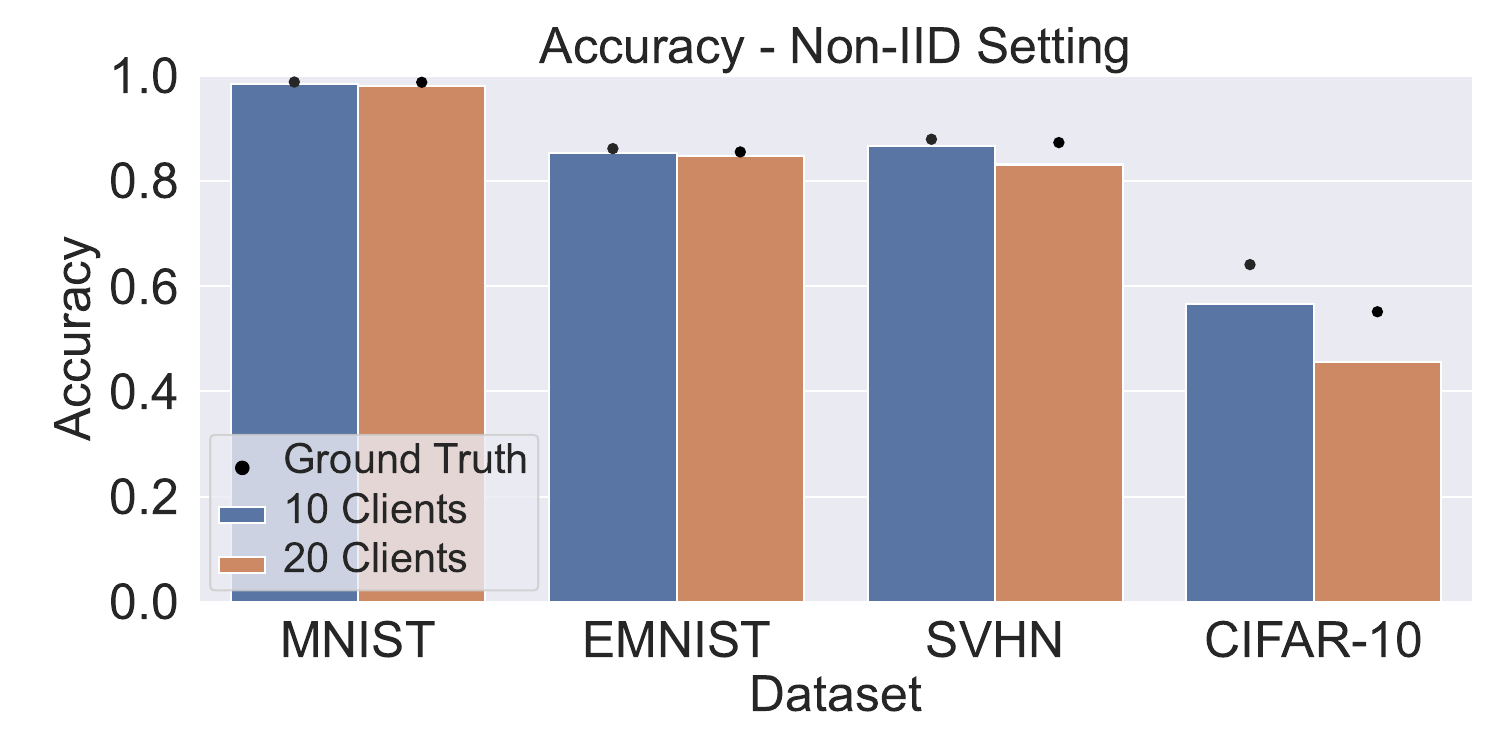} 
    \captionsetup{width=\textwidth}
    \caption{Accuracy}
    \label{fig:noniid-feature-accuracy}
    \end{subfigure}
    \caption{Barplots of learning rate, momentum, and accuracy for the non-iid setting with feature skew of $\beta_{f}=0.02$, using DBSCAN as the $\textsf{Combine}(\cdot)$ strategy on various datasets and experiments. The ground truth (GHO results) is indicated by black dots. Bar colors represent the results of combining the client optimal HPs with the DBSCAN strategy, for variable number of clients ($N$).}
    \label{fig:noniid-feature}
\end{figure*}

\begin{figure*}[t]
\centering
    \begin{subfigure}[b]{.33\textwidth}
    \centering
    \includegraphics[width=\textwidth]{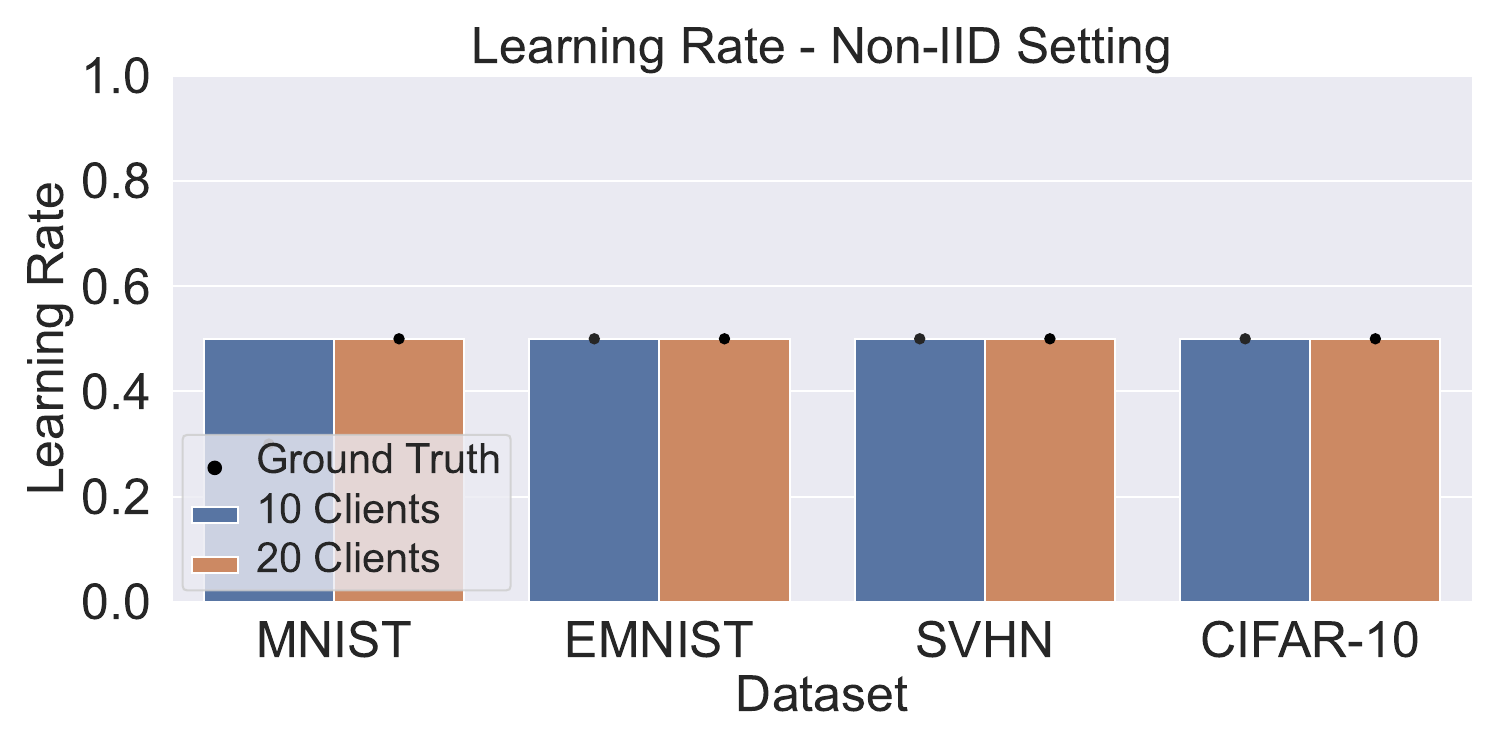}
    \captionsetup{width=\textwidth}
    \caption{Learning Rate}
    \end{subfigure}
    \hfill
    \begin{subfigure}[b]{.33\textwidth}
    \centering
    \includegraphics[width=\textwidth]{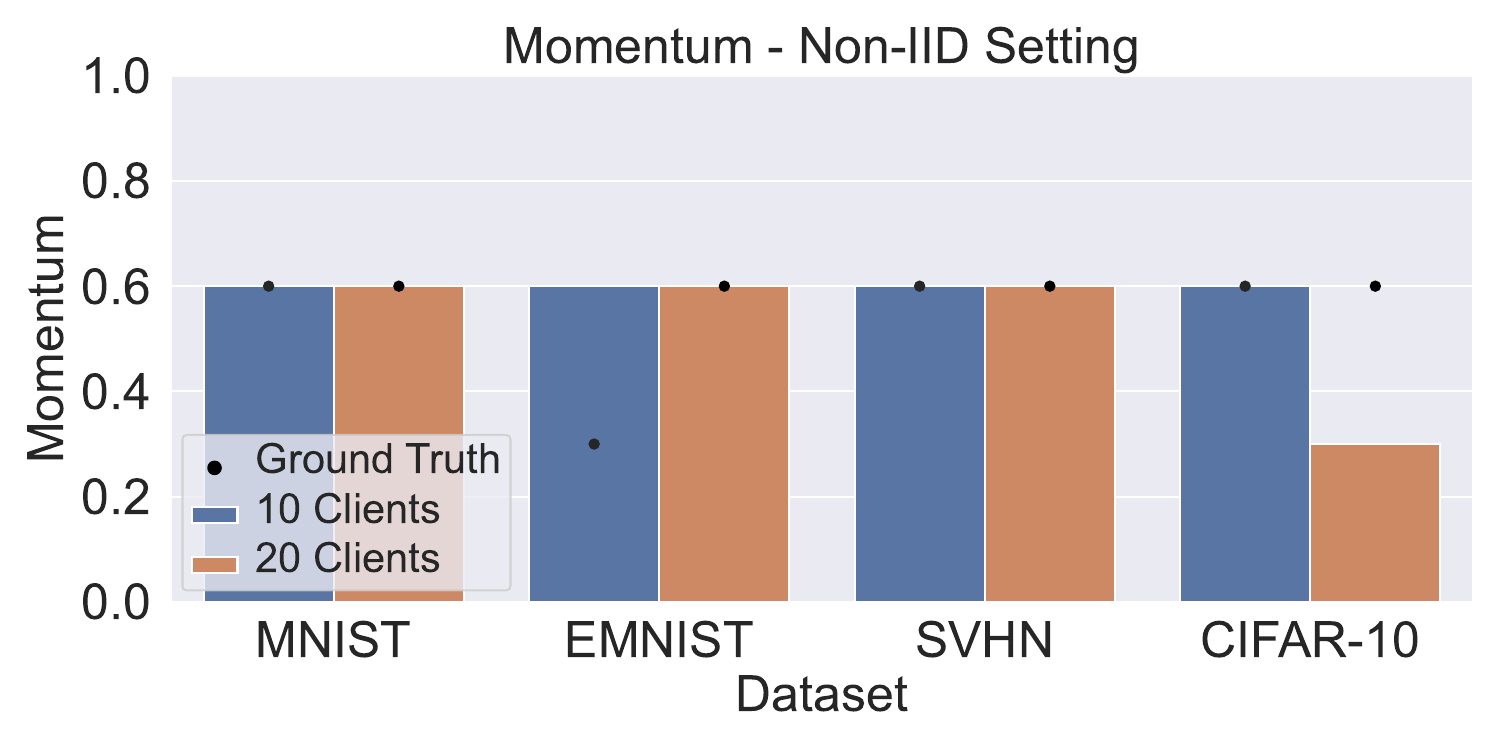}
    \captionsetup{width=\textwidth}
    \caption{Momentum}
    \end{subfigure}
    \hfill
    \begin{subfigure}[b]{.33\textwidth}
    \centering
    \includegraphics[width=\textwidth]{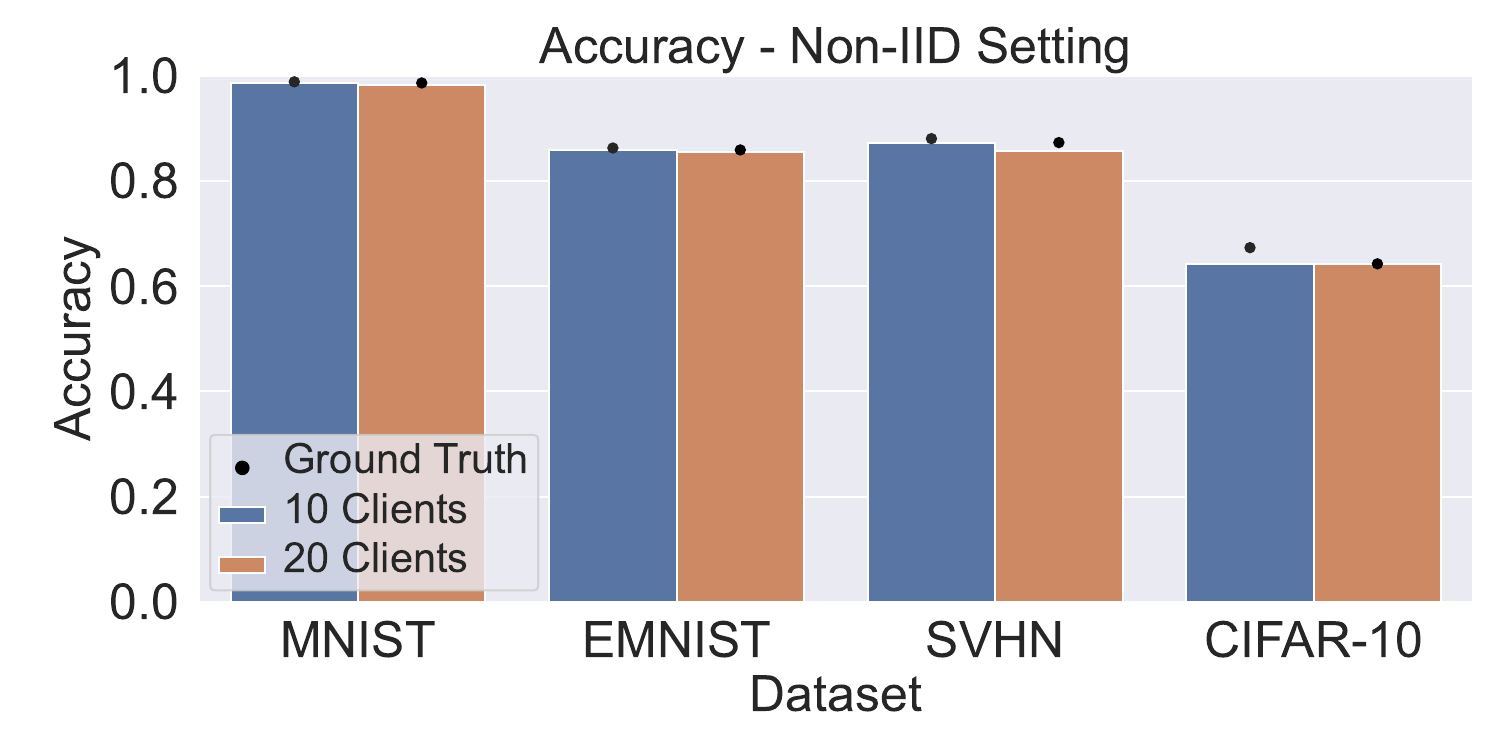} 
    \captionsetup{width=\textwidth}
    \caption{Accuracy}
    \label{fig:noniid-quantity-accuracy}
    \end{subfigure}
    \caption{Barplots of learning rate, momentum, and accuracy for the non-iid setting with quantity skew of $\beta_{q}=2.0$, using DBSCAN as the $\textsf{Combine}(\cdot)$ strategy on various datasets and experiments. The ground truth (GHO results) is indicated by black dots. Bar colors represent the results of combining the client optimal HPs with the DBSCAN strategy, for variable number of clients ($N$).}
    \label{fig:noniid-quantity}
\end{figure*}

Density-based Spatial Clustering of Applications with Noise (DBSCAN) is an algorithm that aims at identifying clusters of high density in a given dataset. DBSCAN is particularly effective at finding clusters with arbitrary shapes. It operates by randomly selecting a point from the dataset and by expanding a cluster around it if the point exhibits sufficient density. In particular, the density is calculated by considering all points within a certain distance $\mathcal{E}$ from the selected point. A cluster is formed if the number of nearby neighboring points is higher than a threshold, i.e., the minimum number of neighboring points $\textsf{MinPts}$. This process is executed iteratively until all points in the dataset have been assigned to a cluster or marked as an outlier. Choosing appropriate values for the $\mathcal{E}$ and $\textsf{MinPts}$ parameters is crucial for the efficient utilization of DBSCAN as these determine the shape and the size of the discovered clusters. DBSCAN has a low number of parameters, which are straightforward to tune. Existing literature has analyzed its performance on multiple settings and rules of thumb have been proposed. For instance, Sander et al.~\cite{sander} suggest that $\textsf{MinPts} = 2 * d$, where $d$ denotes the dimensionality of the data. Accordingly, a k-distance plot for $k = \textsf{MinPts} + 1$, which illustrates the distance of each point to its k-th nearest neighbor, can be used~\cite{Ozkok} to tune $\mathcal{E}$ as the point of a significant change in the plot (i.e., the ``knee" of the plot).

\subsection{Supplementary Figures} \label{sec:additionalFigures}

Figures~\ref{fig:measurement-2},~\ref{fig:noniid-feature}, and~\ref{fig:noniid-quantity}, display additional experimental results from our measurement study (discussed in more detail in Section~\ref{sec:results}).

\end{document}